\documentclass[12pt,a4paper]{article}
\usepackage{amsmath,amsfonts,amsthm,graphicx}
\allowdisplaybreaks[4]
\numberwithin{equation}{section}

\usepackage[body={16cm,23cm}]{geometry}
\newcommand{\cc}{{\mathbf c}}
\newcommand{\cd}{{\mathbf c}^{\dagger}}
\newcommand{\n}{{\mathbf n}}

\newcommand{\Lc}{{\mathcal L}}
\newcommand{\Lf}{{\mathbf L}}
\newcommand{\Lbf}{\overline{\mathbf L}}
\newcommand{\Lo}{\overline{L}}
\newcommand{\Rf}{{\mathbf R}}

\newcommand{\eb}{\overline{e}}
\newcommand{\Ik}{{\mathfrak I}}
\begin{document}
\title{
A note on 
$q$-oscillator realizations of 
 $U_{q}(gl(M|N))$ for 
Baxter $Q$-operators
}
\author{Zengo Tsuboi
\\[8pt]
{\sl
Laboratory of physics of living matter, 
}
 \\
 {\sl
School of biomedicine, 
Far Eastern Federal University, 
}
\\
{\sl
Sukhanova 8, Vladivostok  690950, Russia
}
\\[8pt]
{\sl
Osaka city university advanced mathematical institute,
}
\\
{\sl
3-3-138 Sugimoto, Sumiyoshi-ku Osaka 558-8585, Japan
}
} 
\maketitle
\begin{abstract}
We consider  asymptotic limits of 
q-oscillator (or Heisenberg) realizations of Verma modules over the quantum superalgebra
 $U_{q}(gl(M|N))$, 
and obtain q-oscillator realizations of the 
contracted algebras proposed in \cite{Tsuboi12}. 
Instead of factoring out the invariant subspaces, we 
make reduction on  generators of the q-oscillator algebra, 
which gives a shortcut to the problem. 
Based on this result, 
we obtain explicit q-oscillator representations of 
a Borel subalgebra of the quantum affine superalgebra $U_{q}(\hat{gl}(M|N))$
for Baxter Q-operators. 

%
\end{abstract}
\\[5pt]
Journal reference: Nuclear Physics B 947 (2019) 114747
\\[5pt]
DOI: https://doi.org/10.1016/j.nuclphysb.2019.114747
%
\section{Introduction} 
In the context of quantum integrable systems, 
the Baxter Q-operator \cite{Bax72} is a fundamental object. 
It is known that Baxter Q-operators can be constructed in terms of  
q-oscillator representations of one of the Borel subalgebras of 
quantum affine algebras. This `q-oscillator construction' of the Q-operators 
was proposed by Bazhanov, Lukyanov and Zamolodchikov \cite{BLZ97},  
and developed by many people 
(for instance, see \cite{BHK02,KZ05,BJMST06,Kojima08,BT08,BGKNR10,Tsuboi12,KT14,Mangazeev14,MT15,NR16} and references therein
\footnote{As for the rational ($q=1$) case, see \cite{BLMS10,BFLMS10,FLMS10}. 
There is another approach to Q-operators \cite{D05,DM06,DM10}.}
). 
In particular, Bazhanov, Hibberd and Khoroshkin derived \cite{BHK02} this type of 
q-oscillator representations as asymptotic limits of evaluation 
Verma modules over a Borel subalgebra of $U_{q}(\hat{sl}(3))$.  
Moreover, Hernandez and Jimbo showed \cite{HJ11} that the same type of q-oscillator representations 
can be systematically constructed by taking  
asymptotic limits of Kirillov-Reshetikhin modules over one of the Borel subalgebras of 
 any non-twisted quantum affine algebra. In addition, this approach 
 was further developed \cite{Zhang15,Zhang14} for $U_{q}(\hat{sl}(M|N))$ case. 
 Hernandez and Jimbo's approach is representation theoretically 
 sophisticated, but rather abstract, and thus 
 it is still meaningful to seek another method to obtain explicit 
 q-oscillator realizations, which will be useful for applications to concrete problems. 
 In this paper, we make a proposal on this for $U_{q}(\hat{gl}(M|N))$ case, where 
 we develop, in part, the scheme proposed in our previous paper \cite{Tsuboi12}. 
 In our classification \cite{T09} of the Q-operators, 
there are $2^{M+N}$ kinds of Q-operators for $U_{q}(\hat{gl}(M|N))$, each of 
 which is labeled by a subset $I$ of $\{1,2,\dots, M+N\}$. 
  In the paper \cite{Tsuboi12}, we mainly considered  
 $\mathrm{Card}(I)=0,1,M+N-1,M+N$ cases
 \footnote{We already 
 gave q-oscillator realizations of the diagonal elements of the L-operators for any $I$.}. 
 In this paper, we propose q-oscillator realizations 
 for  $2 \le \mathrm{Card}(I) \le M+N-2 $ case.

In general, the Kirillov-Reshetikhin modules are considered to be  derived from 
 Verma modules based on a procedure, called the BGG-resolution. 
This implies that 
one has to factor out unnecessary invariant subspaces to get the final results 
if one starts from Verma modules \cite{BHK02,NR16}. 
In this paper, we also start from Verma modules, but realize them in terms of 
 the q-oscillator algebra based on the Heisenberg realization (q-difference realization) of 
 $U_{q}(gl(M|N))$ \cite{Kimura96,AOS97} on the 
 flag manifold (for $N=0$ case, \cite{ANO94})
from the very beginning, and then consider 
reduction on generators of the q-oscillator algebra, from which we obtain 
various q-oscillator realizations of $U_{q}(gl(M|N))$ that interpolate 
the full Verma module and the simplest q-oscillator realization, namely,
 the q-Holstein-Primakoff type realization (cf.\ \cite{P98}). 
By taking limits of them, we obtain q-oscillator realizations of contracted algebras
\footnote{A preliminary form of the contracted algebras 
was proposed  in \cite{Bp} for $(M,N)=(3,0)$ case, 
and in \cite{talks} for $(M,N)=(2,1)$ case.}
 $U_{q}(gl(M|N;I))$ for $U_{q}(gl(M|N))$ \cite{Tsuboi12}, and those of 
 the q-super-Yangian $Y_{q}(gl(M|N))$ via an evaluation map. 
 A merit to consider reduction on the q-oscillator algebra lies in the fact that 
 we do not have to factor out invariant subspaces, and thereby are able to take a 
 shortcut to the problem.
We remark that the rational limit ($q \to 1$) of our results reproduce 
 the L-operators for Q-operators associated with $Y(gl(M|N))$ \cite{FLMS10}
 (see \cite{BFLMS10} for $N=0$ case). 
 
 We also remark that the q-oscillator representations of one of the Borel subalgebras 
 of the quantum affine algebra can not be straightforwardly extended to those of 
 the whole quantum affine algebra. 
 The extended representations 
 could be interpreted \cite{Tsuboi12} 
as those of contracted algebras of the original 
 algebra. 
  We will deal with limits of representations of the 
 `whole' \footnote{This `whole' is for the Chevalley generators. 
 In the FRT formulation of the quantum affine algebra, we need only `half' of 
 the algebra  (q-Yangian) since we only consider evaluation representations. 
 In this sense, we may say that we are still dealing with only one of the 
 Borel subalgebras of the quantum 
 affine algebra rather than the whole algebra.} 
  quantum affine superalgebra keeping in mind applications to 
 Q-operators for open boundary spin chains \cite{FS15,BT17}. Note that 
 the generalized q-Onsager algebra \cite{BB10} and the augmented q-Onsager algebra
 \footnote{The higher rank analogue of the augmented q-Onsager algebra 
 has not been fully understood yet (cf. \cite{Tsuboi17}).} \cite{IT,BB12}, 
 which are underlying algebras for open boundary spin chains, 
 are realized by the generators of the whole quantum affine algebra rather than 
 one of the Borel subalgebras.  

The layout of the paper is the following.
In section 2, we review the relevant quantum superalgebras. 
In particular, q-oscillator realizations of $U_{q}(gl(M|N))$ are 
introduced  based on \cite{Kimura96,AOS97}  (and \cite{ANO94}). 
 The contracted algebras $U_{q}(gl(M|N;I))$ for $U_{q}(gl(M|N))$ \cite{Tsuboi12} 
 are quoted as well. 
 Section 3 deals with our main results, where limits of the q-oscillator 
 realizations are taken. 
In section 4, we take the rational limit of the our results 
and make comparison with the rational L-operators for Q-operators \cite{FLMS10}. 
Section 5 is for concluding remarks. In Appendix A, commutation relations of 
$U_{q}(gl(M|N))$ and $U_{q}(gl(M|N;I))$ are summarized in our convention. 
In Appendix B, we transcribe the
 Heisenberg realization of  $U_{q}(gl(M|N))$ in \cite{Kimura96,AOS97}  (and \cite{ANO94}) 
 in terms of the q-oscillator algebra, and review four kinds of variations of them, one of 
 which is used in the main text. 
 Appendix C is a supplement for subsection \ref{q-rel-cont}. 
 Appendix D is a supplement for our previous paper \cite{Tsuboi12}, 
 in which q-Holsten-Primakoff realizations of $U_{q}(gl(M|N))$ 
 are used to rederive the L-operators for Q-operators. 

Throughout this paper, we assume that the deformation parameter $q$ is not 
a root of unity, and 
use the following notation.
\begin{itemize}
\item
$[x]_{q}=(q^{x}-q^{-x})/(q-q^{-1})$ 

\item
${\mathfrak I}=\{1,2,\dots, M+N \}$
\item
$p$:  
the $\mathbb{Z}_{2}$-grading parameter, 
$p(i)=0$ for $i \in {\mathfrak B}$ and 
$p(i)=1$ for $i \in {\mathfrak F}$, 
where ${\mathfrak B}$ is any subset of ${\mathfrak I}$ with 
 $\mathrm{Card}({\mathfrak B})=M$, and 
 ${\mathfrak F}={\mathfrak I}\setminus {\mathfrak B}$.

\item 
$p_{i}=(-1)^{p(i)}$ for $i \in {\mathfrak I}$ 

\item $[\cdot , \cdot ]_{q}$:
q-super-commutator, 
$[X,Y]_{q}=XY-(-1)^{p(X)p(Y)}q YX$,   
$[X,Y]_{1}=[X,Y]$

\item
 $E_{ij}$: the $(M+N) \times (M+N)$ matrix unit with the parity 
$p(E_{ij})=p(i)+p(j) \mod 2$. The $(k,l)$-element of it is $\delta_{i,k} \delta_{j,l}$.  
 
 \item $\theta$: the function defined by 
 $\theta(\text{True})=1$  and  $\theta(\text{False})=0$
 
 \item $\otimes$: the super (graded) tensor product, 
 $(A \otimes B)(C \otimes D)=(-1)^{p(B)p(C)}(AC \otimes BD)$ 
 for homogeneous elements
 
 \item 
 $\n_{i,[b,c]}=\sum_{j=b}^{c}\n_{i,j}$, 
  $\n_{[b,c],i}=\sum_{j=b}^{c}\n_{j,i}$,
  $\n_{I,i}=\sum_{j \in I}\n_{j,i}$, 
  $\n_{i,I}=\sum_{j \in I}\n_{i,j}$, 
  $p_{[b,c]}=\sum_{j=b}^{c}p_{j}$,  
  $p_{I}=\sum_{j \in I}p_{j}$ for $I \subset {\mathfrak I}$.
 \end{itemize}
\section{Quantum superalgebras}
In this section, we review the quantum affine superalgebra $U_{q}(\hat{gl}(M|N))$, 
the quantum finite algebra $U_{q}(gl(M|N))$ and the contracted algebras $U_{q}(gl(M|N;I))$ for it. 
\subsection{The quantum affine superalgebra $U_{q}(\hat{gl}(M|N))$ }
The quantum affine superalgebra $U_{q}(\hat{gl}(M|N))$ \cite{Yamane99} 
(see also \cite{KT94}) is a ${\mathbb Z}_{2}$-graded Hopf algebra 
generated by the generators
\footnote{In this paper, we do not use the degree operator $d$. 
We will only consider  level zero representations. 
The notations $e_{0},f_{0}$ in the 
previous paper \cite{Tsuboi12} correspond to $e_{M+N},f_{M+N}$ 
in this paper.}
 $e_{i},f_{i},k_{i}$, where 
 $i \in {\mathfrak I}$. 
We assign the parity for these generators as 
$p(e_{i})=p(f_{i})=p(i)+p(i+1)  \mod 2 $ and 
$p(k_{i})=0$, where $p(M+N+1)=p(1)$. 
For any $X,Y \in U_{q}(\hat{gl}(M|N))$, 
we define $p(XY)=p(X)+p(Y) \mod 2$. 
For $i,j \in {\mathfrak I}$, the
 defining relations
  of the algebra $U_{q}(\hat{gl}(M|N))$ are 
given by 
\begin{align}
& [k_{i} ,k_{j}]=0, \qquad [k_{i},e_{j}] =(\delta_{ij} -\delta_{i,j+1} )e_{j}, 
\qquad [k_{i},f_{j}] =-(\delta_{ij} -\delta_{i,j+1} )f_{j}, 
\label{efk-0}
\\
&[e_{i},f_{j}]=\delta_{ij} \frac{q^{h_{i}} -q^{-h_{i}} }{q-q^{-1}}, 
\label{efk-1}
\\
& [e_{i},e_{j}]= [f_{i},f_{j}]=0 \quad \text{for} \quad a_{ij}=0, 
 \label{com-ee}
\end{align}
where $h_{i}=p_{i} k_{i} - p_{i+1} k_{i+1}$;    
$(a_{ij})_{1 \le i,j\le M+N}$ is the Cartan matrix 
\begin{align}
a_{ij}=(p_{i}+p_{i+1})\delta_{ij}-
p_{i+1}\delta_{i,j-1}-p_{i}\delta_{i,j+1}.  
 \label{Cartan-mat}
\end{align}
Here  $i,j$ should be interpreted modulo $M+N$: $p_{M+N+1}=p_{1}$, 
$\delta_{i,M+N+1}= \delta_{i,1},
\delta_{i,0}= \delta_{i,M+N}$.
%
%
In addition to the above relations, there are Serre relations 
(see \ \cite{Yamane99}, for more details).  
The algebra also has the
 co-product, anti-poide and co-unit, which will not be used in this paper. 

The Borel subalgebras ${\mathcal B}_{+}$ 
(resp. ${\mathcal B}_{-}$) is generated by 
$e_{i}, k_{i} $ (resp. $f_{i},k_{i}$), where 
$i \in {\mathfrak I}$. 
 For any $c_{i} \in {\mathbb C}$ (multiplied by a unit element), 
the following transformation 
\begin{align}
k_{i} \mapsto k_{i} +p_{i} c_{i} \qquad \text{for} \quad i \in \Ik
\label{shiftgl}
\end{align}
gives the shift automorphism of the Borel subalgebras  $ {\mathcal B}_{+} $ or ${\mathcal B}_{-} $. 


\subsection{The quantum superalgebra $U_{q}(gl(M|N))$}
There is a (finite) quantum superalgebra $U_{q}(gl(M|N))$, which is generated 
by the elements $\{ e_{ij} \}_{i,j \in \Ik}$. We assign the parity 
of these generators as $ p(e_{ij})=p(i)+p(j) \mod 2$. 
Let us introduce the notation: 
$e_{\alpha_{i}}= e_{i,i+1}$,  $e_{-\alpha_{i}}=e_{i+1,i}$ 
for $i \in {\mathfrak I}\setminus \{M+N \}$. 
Then the defining relations of $U_{q}(gl(M|N))$ 
are (cf.\  \cite{KT91}) 
\begin{align}
& [e_{ii},e_{jj}]=0, \quad 
[e_{ii},e_{\pm \alpha_{j}}]=\pm (\delta_{i,j}-\delta_{i,j+1} ) e_{\pm \alpha_{j}}, 
\nonumber
\\ & 
[e_{\alpha_{i}}, e_{-\alpha_{j}}] = 
p_{i} \delta_{ij} 
 \frac{q^{ p_{i}e_{ii}- p_{i+1}e_{i+1,i+1}} - 
q^{- p_{i}e_{ii}+ p_{i+1}e_{i+1,i+1}}}{q-q^{-1}} , 
\nonumber
\\
& [e_{\alpha_{i}},e_{\alpha_{j}}]=
[e_{-\alpha_{i}},e_{-\alpha_{j}}]=0 \quad \text{for} 
\quad |i-j|\ge 2, 
\\
& [e_{\alpha_{i}},[e_{\alpha_{i}},e_{\alpha_{j}}]_{q}]_{q^{-1}}=
[e_{-\alpha_{i}},[e_{-\alpha_{i}},e_{-\alpha_{j}}]_{q^{-1}}]_{q}=0 
\quad \text{for}  \quad |i-j|= 1 \quad \text{and} \quad  p(e_{ \pm \alpha_{i}} ) =0 , 
\nonumber
\\ & [e_{\pm \alpha_{i}},e_{\pm \alpha_{i}}]=0 ,
\nonumber
\\
&  [e_{\alpha_{i}},[e_{\alpha_{i+1}}, [e_{\alpha_{i}}, e_{\alpha_{i-1}} ]_{q^{-1}}]_{q} ] =
[e_{-\alpha_{i}},[e_{-\alpha_{i+1}}, [e_{-\alpha_{i}}, e_{-\alpha_{i-1}} ]_{q}]_{q^{-1}} ] =0 
\quad \text{for}  \quad   p(e_{ \pm \alpha_{i} }) =1 .
\nonumber 
\end{align}
The other elements are defined by 
\begin{align}
\begin{split}
e_{ij}&=[e_{ik},e_{kj}]_{q^{p_{k}}} \qquad 
\text{for} \quad i>k>j, \\
e_{ij}&=[e_{ik},e_{kj}]_{q^{-p_{k}}} \qquad 
\text{for} \quad i<k<j.
\end{split}
\label{eij}
\end{align}
We summarize the relations among these elements in 
 Appendix A. 
There is an evaluation map 
 $\mathsf{ev}_{x}$: 
$U_{q}(\hat{gl}(M|N)) \mapsto U_{q}(gl(M|N))$:  
\begin{align}
\begin{split}
& e_{M+N} \mapsto x q^{-p_{1}e_{11}} e_{M+N,1} q^{-p_{M+N}e_{M+N,M+N}},   \\
& f_{M+N} \mapsto p_{M+N}x^{-1} q^{p_{M+N}e_{M+N,M+N}} e_{1,M+N} q^{p_{1}e_{1,1}}, 
 \\
& e_{i} \mapsto e_{i,i+1}, \qquad 
f_{i} \mapsto p_{i}e_{i+1,i} 
   \qquad \text{for} \quad   i \in \Ik \setminus \{M+N \},
 \\
& 
k_{i} \mapsto e_{ii} \qquad \text{for} \quad   i \in \Ik , 
\end{split} 
\label{eva}
\end{align}
where $x  \in {\mathbb C}\setminus \{0\}$ is a spectral parameter. 
\subsection{q-oscillator realizaiton of $U_{q}(gl(M|N))$}
%
In \cite{AOS97,Kimura96}, a q-difference (Heisenberg) realization of $U_{q}(sl(M|N))$ 
was proposed (see \cite{ANO94} for $U_{q}(sl(M))$ case). 
In this paper, 
we transcribe their results for $U_{q}(gl(M|N))$ case 
in terms of the q-oscillator algebra (the exact relation to their convention is encapsulated in 
Appendix B). 
 
The $q$-oscillator (super)algebra
\footnote{$\cc_{ia}$ in this paper corresponds to $\cc_{ai}$ 
in our  previous paper \cite{Tsuboi12}.} is generated by the 
generators $\{ \cc_{ia},\cd_{ia}, \n_{ia} \}_{i,a \in \Ik, i<a}$,  
whose parities are defined by 
$p(\cc_{ia})=p(\cd_{ia})=p(a)+p(i) \mod 2$, $p(\n_{ia})=0$. 
They obey the following defining relations: 
\begin{align}
\begin{split}
& [\cc_{ia}, \cd_{jb}]_{q^{p_{a}\delta_{ab}\delta_{ij}}}  =\delta_{ab} \delta_{ij} q^{- p_{i} \n_{ia}}, 
\quad 
 [\cc_{ia}, \cd_{jb}]_{q^{-p_{a}\delta_{ab}\delta_{ij}}}  =\delta_{ab} \delta_{ij} q^{p_{i} \n_{ia}} , 
\\[6pt]
&  
[\n_{ia}, \cc_{jb}]=-\delta_{ij}\delta_{ab} \cc_{jb}, \quad 
 [\n_{ia}, \cd_{jb}]=\delta_{ij}\delta_{ab} \cd_{jb}, \quad
[\n_{ia}, \n_{jb}]=[\cc_{ia}, \cc_{jb}]=[\cd_{ia}, \cd_{jb}]=0.
\end{split}
\label{qosc}
\end{align}
From  \eqref{qosc}, 
one can derive useful relations
\footnote{Let us consider the case $p_{i}=-p_{a}$. In this case, $ (\cc_{ia})^{2}=0$ holds. 
Then the relation $0=\cd_{ia}(\cc_{ia})^{2}=[\n_{ia}]_{q}\cc_{ia}$ 
reduces to $q^{\n_{ia}}\cc_{ia}=q^{-\n_{ia}}\cc_{ia}$, 
which is equivalent to $q^{p_{i}\n_{ia}}\cc_{ia}=q^{p_{a}\n_{ia}}\cc_{ia}$. 
Note that this becomes a trivial identity for the case $p_{i}=p_{a}$. 
The relation $\cd_{ia}q^{p_{i}\n_{ia}}=\cd_{ia}q^{p_{a}\n_{ia}}$ can be derived similarly.}
:  
 $ \cc_{ia}\cd_{ia}=[1+ p_{i}p_{a}\n_{ia}]_{q}$, 
$ \cd_{ia}\cc_{ia}=[\n_{ia}]_{q}$, 
$q^{p_{i}\n_{ia}}\cc_{ia}=q^{p_{a}\n_{ia}}\cc_{ia}$ and 
$\cd_{ia}q^{p_{i}\n_{ia}}=\cd_{ia}q^{p_{a}\n_{ia}}$.
The Fock space is spanned by the vectors 
\begin{align} 
| \{ n_{jb} \}_{j,b \in \Ik, j<b} \rangle= 
\overrightarrow{\prod_{j=1}^{M+N-1}} 
\overrightarrow{\prod_{b=j+1}^{M+N}}   (\cd_{jb})^{n_{jb}} |0 \rangle ,  
\label{vecFock}
\end{align}
where $n_{jb} \in \mathbb{Z}_{\ge 0}$ for $p_{j}p_{b}=1$ and 
$n_{jb} \in \{0,1 \}$ for $p_{j}p_{b}=-1$, 
and the vacuum vector is defined by 
\begin{align} 
\n_{ia}|0 \rangle=\cc_{ia}|0\rangle=0 
\quad \text{for all} \quad i, a \in \Ik, \quad  i<a. 
 \label{vacosc}
\end{align}
The action of the generators on $| \{ n_{jb} \} \rangle =| \{ n_{jb} \}_{j,b \in \Ik, j<b} \rangle $ is 
\begin{align}
\begin{split}
\cd_{ia} | \{ n_{jb} \} \rangle &  =
(-1)^{\sum_{k<i} \sum_{k<d} n_{kd}(p(i)+p(a))(p(k)+p(d)) +  \sum_{i<d<a} n_{id}(p(i)+p(a))(p(i)+p(d))}
| \{ n_{jb} + \delta_{ij}\delta_{ab} \} \rangle ,
\\[6pt] 
\cc_{ia} | \{ n_{jb} \} \rangle &  =
(-1)^{\sum_{k<i} \sum_{k<d} n_{kd}(p(i)+p(a))(p(k)+p(d)) +  \sum_{i<d<a} n_{id}(p(i)+p(a))(p(i)+p(d))}
\\[6pt]
& \times 
[1 + (-1)^{p(i)+p(a)} (n_{ia} -1) ]_{q} 
| \{ n_{jb} - \delta_{ij}\delta_{ab} \} \rangle , 
\\[6pt] 
\n_{ia} | \{ n_{jb} \} \rangle &  =n_{ia} | \{ n_{jb}  \} \rangle . 
\end{split}
\label{action-osc}
\end{align}
For $\lambda_{i} \in {\mathbb C}$ ($i \in \Ik$),  
 $U_{q}(gl(M|N))$ is realized by 
\begin{align}
e_{ii} & =\lambda_{i} +\n_{[1,i-1],i} - \n_{i,[i+1,M+N]}
 \quad \text{for} \quad  i \in \Ik, 
\nonumber  \\[6pt]
e_{i,i+1} &=  
\sum_{k=1}^{i-1}
 \cd_{ki} \cc_{k,i+1}
\nonumber  \\
& \quad \times 
q^{-p_{i}\lambda_{i}+ p_{i+1} \lambda_{i+1} -
p_{i} \n_{[k+1,i-1],i} + p_{i+1} \n_{[k+1,i],i+1}  +
 p_{i} \n_{i,[i+1,M+N]} - p_{i+1} \n_{i+1,[i+2,M+N]}  } 
\nonumber  \\
 &
 \quad 
+
p_{i}\cc_{i,i+1} 
\left[
p_{i}\lambda_{i}- p_{i+1} \lambda_{i+1}- p_{i} \n_{i,[i+1,M+N]} +
 p_{i+1} \n_{i+1,[i+2,M+N]}  +p_{i}
\right]_{q}
\nonumber  \\
& \quad 
 -p_{i}
\sum_{k=i+2}^{M+N}  p_{k}
\cc_{ik} \cd_{i+1,k} q^{p_{i}\lambda_{i}-p_{i+1}\lambda_{i+1} 
- p_{i}\n_{i,[k,M+N]} +p_{i+1}\n_{i+1,[k,M+N]} +p_{i}+p_{i+1} },
\label{real44}
\\[6pt]
e_{i+1,i} &=  \cd_{i,i+1} q^{ p_{i}\n_{[1,i-1],i} - p_{i+1}\n_{[1,i-1],i+1}  } 
+
\sum_{k=1}^{i-1}  \cd_{k,i+1} \cc_{ki} q^{ p_{i}\n_{[1,k-1],i} - p_{i+1}\n_{[1,k-1],i+1} } 
\nonumber \\
 & \qquad \text{for}  \quad  i \in \Ik \setminus \{M+N \} .
 \nonumber 
\end{align}
In principal, one can  recursively calculate all the generators $e_{ij}$ for $|i-j|\ge 2$ based on 
the relations \eqref{eij}. However, their general expressions are  very involved. 
Fortunately, $e_{i1}$ is tractable and has a simple expression
 (cf. \cite{ANO94} for $M=0$ case):  
\begin{align}
e_{i1}&= \cd_{1i}  q^{ - p_{1}\n_{1,[2,i-1]}} 
 \quad \text{for} \quad   i \in \Ik \setminus \{1 \}.
 \label{enj-44}
\end{align}
On the Fock space, \eqref{real44} realizes a highest weight representation
\footnote{According to \cite{ANO94}, \eqref{real1} 
(which can be transformed to \eqref{real44}) gives 
a Verma module at least for $N=0$ case.  In fact, the action of the generators \eqref{real44} on 
the vector $ | \{ n_{jb}  \} \rangle$ \eqref{vecFock} 
for $N=0$ 
coincides with the one given by eqs.\ (4.3)-(4.6) in \cite{NR16} 
for $N=0$ case under the transformation 
$q \to q^{-1}$. 
Moreover, $\pi_{\lambda}$ is expected to be a Verma module of $U_{q}(gl(M|N))$
for any  $M,N$ 
since the Verma module has a PBW basis in almost the same form as 
\eqref{vecFock} (if $\cd_{jb}$ is replaced by $e_{bj}$) 
[we thank the referee for this comment].
} 
 $\pi_{\lambda}$  with the highest
 weight $\lambda =(\lambda_{1}, \lambda_{2}, \dots, \lambda_{M+N})$ 
and the highest weight vector $ |0 \rangle $ satisfying
\footnote{More generally, $e_{jk} |\lambda \rangle =0$ for  $j<k$ 
follows from \eqref{eij}.} 
\begin{align}
\begin{split}
&e_{ii} |0 \rangle =\lambda_{i} |0 \rangle 
\quad \text{for}  \quad i \in {\mathfrak I},
%
\qquad 
e_{\alpha_{j}} |0 \rangle =0
 \quad \text{for}  \quad j \in {\mathfrak I}\setminus \{M+N\}.
 \end{split}
 \label{hwvglmn}
\end{align}
The composition $\pi_{\lambda } \circ \mathsf{ev}_{x}$ 
gives an evaluation representation of $U_{q}(\hat{gl}(M|N))$. 
Let us consider reduction of the 
q-oscillator algebra in \eqref{real44}. 
Fix parameters $a \in \{0,1,\dots, M+N\}$ and $\mu \in {\mathbb C}$, and define a set by 
$I=\{a+1,a+2,\dots, M+N \}$. 
We find that 
\eqref{real44} still realizes $U_{q}(gl(M|N))$ 
even if we apply the following replacement:
\begin{align}
\cc_{ij} \mapsto 0 , 
\quad 
\cd_{ij} \mapsto 0,
\quad 
\n_{ij} \mapsto 0 ,
\quad 
\lambda_{i} \mapsto p_{i}\mu 
\quad \text{for} \quad i,j \in I. 
 \label{rel44-res}
\end{align}
This fact is remarked in \cite{ANO94} 
for $N=0$, $a=1$, $\mu=0$ case, where \eqref{real44} 
reduces to a q-analogue of the Holstein-Primakoff realization (cf. \cite{P98}). 
One can easily calculate all the generators $e_{ij}$ for $a=1$ case through \eqref{eij}.
\begin{align}
e_{11}&=\lambda_{1}-\n_{1,I}, 
\qquad e_{ii}=p_{i}\mu+\n_{1i} \quad \text{for} \quad i \in I,
\nonumber
\\[6pt]
e_{1j}&=p_{1}\cc_{1j}[p_{1}\lambda_{1}-\mu-p_{1}\n_{1,[2,M+N]}+p_{1}]_{q}
 q^{p_{1}\n_{1,[2,j-1]}} \quad \text{for} \quad j \in I,
 \nonumber
 \\[6pt]
 e_{ij}&=\cd_{1i}\cc_{1j}
 q^{p_{1}\n_{1,[i+1,j-1]}} \quad \text{for} \quad 2\le i<j,
 \label{eija=1}
  \\[6pt]
e_{i1}&= \cd_{1i}  q^{ - p_{1}\n_{1,[2,i-1]}} 
 \quad \text{for} \quad  i \in I.
 \nonumber
 \\[6pt]
 e_{ij}&=\cd_{1i}\cc_{1j}
 q^{-p_{1}\n_{1,[j+1,i-1]}} \quad \text{for} \quad 2\le j<i.
 \nonumber
\end{align} 
\subsection{FRT realization of  $Y_{q}(gl(M|N))$}
The quantum affine superalgebra $U_{q}(\hat{gl}(M|N))$ (and its subalgebra $U_{q}(gl(M|N))$)  
has another realization, called FRT realization \cite{Faddeev:1987ih} 
(see also, \cite{Zhang97,FM01}), 
based on the Yang-Baxter relation. 
One of the merits of this realization is that all the relations among the generators 
can be expressed in a unified manner independent of $M,N$ 
and the grading parameters $p(i)$. 
While in the realization based on the Chevalley generators, which we mentioned 
in  subsections 2.1 and 2.2, 
the form of the Serre type relations depends sensitively on $M,N$ and $p(i)$, 
and it is rather cumbersome to write 
down all the necessary relations without omission. 
In this sense, the FRT realization, which we are going to explain, supersedes the previous ones. 

The quantum  affine superalgebra $U_{q}(\hat{gl}(M|N))$ has 
 a subalgebra called $q$-super-Yangian  $Y_{q}(gl(M|N))$. 
It is generated by the generators 
$\{\Lc_{ij}^{(n)}|i,j\in \Ik, n \in \mathbb{Z}_{\ge 0} \}$ obeying 
the Yang-Baxter relation
\footnote{We will use the notation $A^{12}=\sum_{i}a_{i}\otimes b_{i} \otimes 1$, 
$A^{13}=\sum_{i}a_{i} \otimes 1 \otimes b_{i}$, 
$A^{23}=\sum_{i} 1 \otimes a_{i}\otimes b_{i}$ 
for  an element of the form $A=\sum_{i}a_{i}\otimes b_{i}$.}
\begin{align}
& \Rf^{23}(xy^{-1})\Lc^{13}(y)\Lc^{12}(x)=
\Lc^{12}(x)\Lc^{13}(y)\Rf^{23}(xy^{-1}),
\label{rlla3}
\\[6pt]
&\Lc(x)=\sum_{i,j=1}^{M+N} \Lc_{ij}(x) \otimes E_{ij}, 
\qquad 
\Lc_{ij}(x)=\sum_{n=0}^{\infty} \Lc_{ij}^{(n)} x^{-n},
\nonumber \\[6pt]
& \Lc_{ij}^{(0)}=0 
\quad \text{for} \quad 1 \le i<j \le M+N ,
 \label{rlla1}
\\[6pt]
& \Rf (x) = \Rf  -x \, \overline{\Rf}, 
 \label{PS-R}
\\
& \Rf = \sum_{i=1}^{M+N} q^{p_{i}} E_{ii} \otimes E_{ii} +
\sum_{i \ne j}  E_{ii} \otimes E_{jj} +
(q-q^{-1}) 
\sum_{i<j} p_{j} E_{ij} \otimes E_{ji},
\nonumber  
\\
& \overline{\Rf} = \sum_{i=1}^{M+N} q^{-p_{i}} E_{ii} \otimes E_{ii} +
\sum_{i \ne j}  E_{ii} \otimes E_{jj} -
(q-q^{-1}) 
\sum_{i>j} p_{j} E_{ij} \otimes E_{ji} . 
\nonumber 
\end{align}
where $x,y \in {\mathbb C}$. 
The parity of the generator is defined by  $p(\Lc^{(n)}_{ij})=
p(\overline{\Lc}^{(n)}_{ij})=p(i)+p(j) \mod 2$.
Here we assume that the elements $\{\Lc_{ii}^{(0)}|i\in \Ik \}$ are invertible. 
$\Rf(x)$ is 
the R-matrix for the Perk-Schultz model \cite{Perk:1981nb}  
 (see \cite{Cherednik80} for $N=0$ case). 

For any $c \in \mathbb{C} \setminus \{0\}$, 
\begin{align}
\Lc(x) \mapsto \Lc(cx), 
 \label{shiftconst}
\end{align}
gives an automorphism of $Y_{q}(gl(M|N))$. 
Note that the following transformation (multiplication of diagonal matrices in the second space) 
\begin{multline}
  \Lc(x) \mapsto  ( 1 \otimes \mathcal{H}_{L}) \Lc(x)  (1 \otimes \mathcal{H}_{R}), 
 \\
 \mathcal{H}_{L}=\sum_{i} \mathcal{H}_{L}^{(i)} E_{ii}, 
\quad \mathcal{H}_{R}=\sum_{i} \mathcal{H}_{R}^{(i)} E_{ii}, 
\quad  \mathcal{H}_{L}^{(i)} ,\mathcal{H}_{R}^{(i)}  \in {\mathbb C} \setminus \{ 0\}
 \label{shiftL} 
\end{multline}
keeps the relations \eqref{rlla1}  and \eqref{rlla3} unchanged.

\subsection{FRT realization of $U_{q}(gl(M|N))$}
The quantum affine superalgebra  $U_{q}(\hat{gl}(M|N))$ has a finite subalgebra 
  $U_{q}(gl(M|N))$. It is generated by the generators 
  $\{\Lf_{ij},\Lbf_{ij},|i,j \in \Ik \}$
  obeying the relations 
\begin{align}
& L_{ij}=\overline{L}_{ji}=0, 
\quad \text{for} \quad 1 \le i<j \le M+N 
\label{rll0}
\\
& L_{ii} \overline{L}_{ii}=\overline{L}_{ii}L_{ii}=1
\quad \text{for} \quad i \in \Ik, 
\label{rll1} \\
& \Rf^{23}\Lf^{13}\Lf^{12}=\Lf^{12}\Lf^{13}\Rf^{23}, 
 \label{rll2}
\\
& \Rf^{23}\Lbf^{13}\Lbf^{12}=\Lbf^{12}\Lbf^{13}\Rf^{23},
\label{rll3}
\\ 
& \Rf^{23}\Lf^{13} \Lbf^{12}=\Lbf^{12} \Lf^{13}\Rf^{23}, 
\label{rll4}
\\[6pt]
& \Lf =\sum_{j,k=1}^{M+N} L_{kj}  \otimes E_{kj}, 
\qquad 
\Lbf =\sum_{j,k=1}^{M+N} \overline{L}_{kj}  \otimes E_{kj},
\nonumber 
\end{align}
where  the parity of the generators is defined by $p(\Lf_{ij})=
p(\Lbf_{ij})=p(i)+p(j) \mod 2$.  
The coefficients are related to the generators \eqref{eij} 
as (cf.\ \cite{Zhang92}) 
\begin{align}
& L_{ii}=q^{p_{i}e_{ii}}, 
 \label{maprll1}
\qquad 
\overline{L}_{ii}=q^{p_{i}\overline{e}_{ii}}, 
\\
& L_{ij}=p_{i}(q-q^{-1})e_{ji} q^{p_{j} e_{jj}} 
\quad \text{for} \quad i >j, \\
& \overline{L}_{ij}=-p_{i}(q-q^{-1})q^{-p_{i} e_{ii}}e_{ji}  
\quad \text{for} \quad i <j ,
\label{maprll3}
\end{align}
where $\overline{e}_{ii}=-e_{ii}$. 
There is an evaluation map from $Y_{q}(gl(M|N))$  
to $U_{q}(gl(M|N))$ such that
\begin{align}
\Lc (x)& \mapsto \Lf (x)=\Lf -\Lbf x^{-1}. \label{rll-ev1} 
\end{align}
The L-operator $\Lf(x)$ satisfies the following Yang-Baxter relation, 
which is the image of \eqref{rlla3} under this map \eqref{rll-ev1}.
\begin{align}
& \Rf^{23}(xy^{-1})\Lf^{13}(y)\Lf^{12}(x)=
\Lf^{12}(x)\Lf^{13}(y)\Rf^{23}(xy^{-1}).
\label{YB-L}
\end{align}
We will repeatedly use the transformation \eqref{shiftL}, 
which preserves the Yang-Baxter relation \eqref{YB-L} 
 under the evaluation map \eqref{rll-ev1}. 
\subsection{Contraction of $U_{q}(gl(M|N))$}
Let us take a subset $I$ of the set ${\mathfrak I}$ and 
 its complement set 
$\overline{I}:={\mathfrak I} \setminus I$. 
There are $2^{M+N}$ choices of the subsets in this case. 
Corresponding to the set $I$, we consider $2^{M+N}$ kinds of representations of the q-superYangian. 
For this purpose, 
we consider $2^{M+N}$ kinds of  contractions of  $U_{q}(gl(M|N))$. 
At first, we modify the condition \eqref{rll1} and define a contracted algebra as follows. 

The contracted algebra 
$\tilde{U}_{q}(gl(M|N;I))$ is an associative algebra over ${\mathbb C}$ with 
a unit element $1$ and generators $L_{ij}, \Lo_{ij}$ obeying 
 the relations 
\eqref{rll0}, \eqref{rll2}-\eqref{rll4} and 
\begin{align}
& L_{ii} \Lo_{ii}=\Lo_{ii}L_{ii}=1
\quad \text{for} \quad i \in I, 
 \label{red1} 
\\
& \Lo_{ii}=0 \qquad \text{for} \quad i \in \overline{I} . 
\label{red2} 
\end{align}
In addition, we assume the existence of an inverse element $L^{-1}_{ii} $ of $L_{ii}$ for any $i \in \Ik $.
\begin{align}
L_{ii} L^{-1}_{ii} =L^{-1}_{ii} L_{ii} =1.
\label{Linv}
\end{align}

We remark that $L^{-1}_{ii} $ coincides with $\Lo_{ii}$ only for $i \in I$. 
Then we obtain $2^{M+N}$ kinds of algebraic solutions of the graded 
Yang-Baxter equation through the map \eqref{rll-ev1}. 
In addition to the contraction \eqref{red2}, 
we introduce the following subsidiary contraction and define a contracted algebra which 
is smaller than $\tilde{U}_{q}(gl(M|N;I))$. 

Suppose the set $I$ has the form 
$I=\{k+1,k+2,\dots, k+n \}$ for some 
$k \in {\mathbb Z}_{\ge 0}, n \in {\mathbb Z}_{>0}$, then the contracted algebra $U_{q}(gl(M|N;I))$ 
\cite{Tsuboi12} is defined by adding the following relations to $\tilde{U}_{q}(gl(M|N;I))$. 
\begin{align}
L_{ij} &= 0 \qquad \text{for} 
 \quad k+n < i \le M+N  \quad \text{and} \quad  1 \le  j \le k, 
\label{red30} \\
\Lo_{ij} &= 0 \qquad \text{for} \quad 1<i<j \le k \quad \text{or} \quad k+n < i<j \le M+N .  \label{red3}
\end{align}
The contracted algebras can be realized in terms of the generators $e_{ij}$. 
They are related to the non-zero elements $L_{ij},\overline{L}_{ij}$ 
through \eqref{maprll1}-\eqref{maprll3}. 
The conditions corresponding to \eqref{red1}-\eqref{Linv} 
are given by 
\begin{align}
&
q^{p_{i}\overline{e}_{ii}} = 0 \quad \text{for} \quad 
 i \in \overline{I}, 
\qquad 
 \overline{e}_{ii} = -e_{ii} \quad  \text{for} \quad  i \in I.
\label{red300e}
\end{align} 
The conditions corresponding to \eqref{red30} and \eqref{red3} 
are given by 
\begin{align}
\begin{split}
e_{ji} = 0 \qquad & \text{for} 
 \quad k+n < i \le M+N  \quad \text{and} \quad  1 \le  j \le k, \quad \text{or} 
\\
 &  \qquad 1<i<j \le k , \quad \text{or} \quad k+n < i<j \le M+N .  
\end{split}
\label{red30e} 
\end{align} 
In the main text, we will focus
\footnote{We expect that the other cases can be obtained from this case by using automorphisms 
of $U_{q}(gl(M|N))$ or $U_{q}(\hat{gl}(M|N))$ taking note on 
the fact that they are no longer automorphisms of the contracted algebras. 
This remains to be clarified.}
 on the case $k=M+N-n$. 
We remark that  the contracted algebra $U_{q}(gl(3|0;I))$
 for $|I|=1,2$ in terms of the generators $e_{ij}$ 
was proposed by Bazhanov and Khoroshkin \cite{Bp} (see, Appendix A). 
The case $U_{q}(gl(2|1;I))$ was also proposed in \cite{talks}. 
We also note that the q-oscillator algebra can be obtained from a contraction procedure 
of the quantum algebra $U_{q}(sl(2))$ \cite{CK90}.

\subsection{Representations of $Y_{q}(gl(M|N))$}
%
Then combining \eqref{eij}, \eqref{real44}, \eqref{rll-ev1} and \eqref{maprll1}-\eqref{maprll3}, 
we obtain a q-oscillator realization of $Y_{q}(gl(M|N))$. In particular, on the Fock space, this 
gives a highest weight representation with the highest weight $ |0 \rangle$ obeying 
\begin{align}
&\Lc_{ii}(x) |0 \rangle =(q^{p_{i}\lambda_{i}} - x^{-1} q^{- p_{i} \lambda_{i}} )|0 \rangle 
\quad \text{for} \quad  i \in {\mathfrak I}, 
\label{weightl} \\[6pt]
&\Lc_{ij}(x) |0 \rangle =0 
 \quad 
\text{for} \quad  i>j, \quad  i,j \in {\mathfrak I}.
\end{align}
The map \eqref{rll-ev1} also 
gives  an evaluation map from $Y_{q}(gl(M|N))$  
to $U_{q}(gl(M|N;I))$ or $\tilde{U}_{q}(gl(M|N;I))$ if 
the matrix elements of $  \Lf$ and $  \Lbf$ are replaced by 
the ones for 
the corresponding contracted algebra. 
\section{Asymptotic representations of $Y_{q}(gl(M|N))$}
In this section, we will consider asymptotic representations of $Y_{q}(gl(M|N))$. 
\subsection{General strategy}
We will combine the transformations \eqref{shiftconst} and \eqref{shiftL}, 
which preserve the Yang-Baxter relation \eqref{rlla3} under \eqref{rll-ev1}, 
namely \eqref{YB-L}, 
and consider limits of the L-operator. This realizes the contracted algebra and 
asymptotic representations of the q-super-Yangian on the Fock space. 
We will also make reductions on generators of the q-oscillator algebra 
 in order to remove the parts which do not have essential contribution 
 on the action on the Fock space.   

We consider the case $I=\{a+1,a+2,\dots,M+N\}$, $\overline{I}={\mathfrak I} \setminus I$. 
In components, 
$\tilde{\mathbf L}(x)=
{\mathbf L}(x)(1 \otimes q^{-\sum_{i \in \overline{I}}p_{i}\lambda_{i} E_{ii}})$ can be written 
as 
\begin{align}
\tilde{L}_{ij}=q^{-p_{j}\lambda_{j}\theta(j \in \overline{I})}L_{ij}, 
\qquad 
\tilde{\overline{L}}_{ij}=q^{-p_{j}\lambda_{j}\theta(j \in \overline{I})}
\overline{L}_{ij},
\label{renoL22}
\end{align} 
where 
$\tilde{\mathbf L}(x)=\tilde{\mathbf L}-
x^{-1}\tilde{\overline{\mathbf L}}=
\sum_{i,j \in {\mathfrak I}}(\tilde{L}_{ij}-x^{-1}\tilde{\overline{L}}_{ij})\otimes E_{ij}$. 
We can translate this  through \eqref{maprll1}-\eqref{maprll3} in the form 
\begin{align}
\begin{split}
&\tilde{e}_{ii}=e_{ii} -\lambda_{i} \theta(i \in \overline{I}),
\quad
q^{p_{i}\tilde{\overline{e}}_{ii}}=
 q^{p_{i}\overline{e}_{ii} -p_{i}\lambda_{i} \theta(i \in \overline{I})}
=q^{-p_{i}\tilde{e}_{ii} -2p_{i}\lambda_{i} \theta(i \in \overline{I})},
\\ 
&\tilde{e}_{ij}=e_{ij} \quad \text{for} \quad i<j, 
\\ 
&\tilde{e}_{ij}=
q^{-p_{i}\lambda_{i} \theta(i \in \overline{I})-p_{j}\lambda_{j} \theta(j \in \overline{I})}e_{ij} \quad \text{for} \quad i>j. 
\end{split}
\label{renoGen22}
\end{align}
where $\overline{e}_{ii}=-e_{ii}$, and the symbol $\tilde{} $ is assigned to each element in \eqref{maprll1}-\eqref{maprll3}.
Then we find that 
\eqref{renoL22} with \eqref{renoGen22} and \eqref{real44} realize $U_{q}(gl(M|N;I))$ in 
 the limit
 \footnote{We also need a fine tune on the normalization of the generators of the q-oscillator algebra.} 
\begin{align}
|\lambda_{i}| \to \infty \quad 
 \text{for all} \quad i \in \overline{I} \quad  \text{under the condition} \quad 
q^{-p_{i}\lambda_{i}+p_{i+1}\lambda_{i+1}} \to 0. 
\label{qlim}
\end{align}
Here we assume that $q$ is a constant parameter with the condition $|q|\ne 1$. 
In particular, $q^{-p_{i}\lambda_{i}} \to 0$ holds for any $i\in \overline{I}$.  
This type of limit for evaluation Verma modules over 
 ${\mathcal B}_{+}$ for $M=3,N=0$, $a=2$ case and $M>3$, $N=0$, $a=M-1$ case was 
considered in \cite{BHK02} and \cite{NR16}, respectively. 
Now, on the Fock space, the evaluation map
\footnote{in the sense ${\mathcal L}(x) \mapsto \lim \tilde{\mathbf L}(x)$}
 \eqref{rll-ev1} gives a highest weight representation of 
$Y_{q}(gl(M|N))$ with the highest weight $|0 \rangle $ obeying 
\begin{align}
\begin{split} 
&{\mathcal L}_{ii}(x) |0 \rangle =|0 \rangle  
\quad \text{for}  \quad i \in \overline{I}, 
\quad 
{\mathcal L}_{ii}(x) |0 \rangle = (q^{p_{i}\lambda_{i}}-x^{-1}q^{-p_{i}\lambda_{i}})  |0 \rangle  
\quad \text{for} \quad  i \in I,
\\[6pt]
&{\mathcal L}_{ij}(x) |0 \rangle =0
 \quad 
\text{for} \quad  i>j, \quad i,j \in {\mathfrak I}. 
\end{split}
\label{HWcont}
\end{align}
As a variant 
\footnote{
The other option is to consider 
$\tilde{\mathbf L}(x)={\mathbf L}(xq^{-2m})(1 \otimes q^{-m \sum_{i \in I} E_{ii}}) $ 
[cf. eq.\ (3.79) in \cite{Tsuboi12}]. In components, this can be written 
as 
\begin{align}
\tilde{L}_{ij}=q^{-m\theta(j \in I)}L_{ij}, 
\qquad 
\tilde{\overline{L}}_{ij}=
q^{m(2-\theta(j \in I))}\overline{L}_{ij}.
 \label{renoL1}
\end{align}
We can translate this through \eqref{maprll1}-\eqref{maprll3} in the form
\begin{align}
\begin{split}
&\tilde{e}_{i,i}=e_{i,i} -p_{i}m \theta(i \in I),
\qquad 
q^{\tilde{\overline{e}}_{i,i}}=
q^{ \overline{e}_{i,i} +p_{i}m (2-\theta(i \in I))},
\\ 
&\tilde{e}_{i,j}=e_{i,j} \quad \text{for} \quad i<j, 
\qquad 
\tilde{e}_{i,j}=
q^{m(2-\theta(i \in I)-\theta(j \in I))}e_{i,j} \quad \text{for} \quad i>j.
\end{split}
\label{renoGen1}
\end{align}
(In eq.(3.25) in \cite{Tsuboi12}, 
we did not interpret the factor $q^{-p_{i}e_{ii}}$ 
as $q^{p_{i}\overline{e}_{ii}}$. If we did it, we would have 
obtained $\tilde{e}_{i,j}=
q^{m(\theta(j \in I)-\theta(i \in I))}e_{i,j}$ for $ i>j$.) 
%
Then for the parameters be set as 
\begin{align}
\lambda_{i} \to p_{i} m +\lambda_{i}  \quad \text{for} \quad   i \in I,  \quad 
\text{and} \quad  \lambda_{i} \to 0  \quad \text{for} \quad  i \in \overline{I} ,
\label{highestweight1}
\end{align}
 \eqref{renoL1} with \eqref{renoGen1} and  \eqref{real44} 
 realize $U_{q}(gl(M|N;I))$ in the limit $q^{m} \to 0$.  
 See Appendix D. 
}
of the above, we can consider 
the case 
\begin{align}
\lambda_{i} =p_{i} m   \quad \text{for} \quad   i \in \overline{I},
\label{highestweight2}
\end{align}
and take the limit $|m| \to \infty $ under the condition $q^{-m} \to 0 $. This also realizes $U_{q}(gl(M|N;I))$. 
We remark that the above two types of limits give the same result {\em after} reductions on 
 generators of the q-oscillator algebra.
\subsection{q-oscillator realization of contracted algebras}
\label{q-rel-cont}
Now we demonstrate the general strategy   
based on the q-oscillator realization \eqref{real44}. 
We consider the case $I=\{a+1,a+2,\dots, M+N\}$, $\overline{I}={\mathfrak I} \setminus I$. 
Let us apply the following automorphism of the q-oscillator algebra to
 \eqref{real44} and \eqref{enj-44}.
\begin{align}
\cc_{ij} \mapsto q^{-p_{i}\lambda_{i}\theta(i\in \overline{I})+
p_{j}\lambda_{j}\theta(j\in \overline{I})}\cc_{ij},
\quad 
\cd_{ij} \mapsto q^{p_{i}\lambda_{i}\theta(i\in \overline{I})-
p_{j}\lambda_{j}\theta(j\in \overline{I})}\cd_{ij},
\quad 
\n_{ij} \mapsto \n_{ij}. 
 \label{cijreo}
\end{align}
Then in the limit \eqref{qlim}, \eqref{renoGen22} reduces to 
\begin{align}
e_{ii} & =\lambda_{i}\theta(i \in I) +\n_{[1,i-1],i} - \n_{i,[i+1,M+N]}
, \quad 
q^{p_{i}\overline{e}_{ii}}=\theta(i\in I)q^{-p_{i}e_{ii}}
 \quad \text{for} \quad  i \in{ \mathfrak I},
 \nonumber
\\[6pt]
e_{i,i+1} &=  
p_{i}(q-q^{-1})^{-1}\cc_{i,i+1} 
q^{- p_{i+1} \lambda_{i+1}\theta(i+1 \in I) - p_{i} \n_{i,[i+1,M+N]} +
 p_{i+1} \n_{i+1,[i+2,M+N]}  +p_{i}
}
 \nonumber \\
& \quad 
 -p_{i}
\sum_{k=i+2}^{M+N}  p_{k}
\cc_{ik} \cd_{i+1,k} q^{-p_{i+1}\lambda_{i+1} \theta(i+1 \in I) 
- p_{i}\n_{i,[k,M+N]} +p_{i+1}\n_{i+1,[k,M+N]} +p_{i}+p_{i+1} }
\nonumber \\
 & \qquad \text{for} \quad i \in \overline{I},
\nonumber \\[6pt]
e_{i,i+1} &=  
\sum_{k=1}^{i-1}
 \cd_{ki} \cc_{k,i+1}
\nonumber \\
& \quad \times 
q^{-p_{i}\lambda_{i}+ p_{i+1} \lambda_{i+1} -
p_{i} \n_{[k+1,i-1],i} + p_{i+1} \n_{[k+1,i],i+1}  +
 p_{i} \n_{i,[i+1,M+N]} - p_{i+1} \n_{i+1,[i+2,M+N]}  } 
 \nonumber \\
 &
 \quad 
+
p_{i}\cc_{i,i+1} 
\left[
p_{i}\lambda_{i}- p_{i+1} \lambda_{i+1}- p_{i} \n_{i,[i+1,M+N]} +
 p_{i+1} \n_{i+1,[i+2,M+N]}  +p_{i}
\right]_{q}
\label{real44lim}
 \\
& \quad 
 -p_{i}
\sum_{k=i+2}^{M+N}  p_{k}
\cc_{ik} \cd_{i+1,k} q^{p_{i}\lambda_{i}-p_{i+1}\lambda_{i+1} 
- p_{i}\n_{i,[k,M+N]} +p_{i+1}\n_{i+1,[k,M+N]} +p_{i}+p_{i+1} },
\nonumber \\
 & \qquad \text{for} \quad i,i+1 \in I,
 \nonumber \\[6pt]
e_{i+1,i} &= 0 \qquad \text{for} \quad i,i+1 \in \overline{I},
\nonumber \\[6pt]
e_{i+1,i} &=   
\cd_{i,i+1} q^{ p_{i}\n_{[1,i-1],i} - p_{i+1}\n_{[1,i-1],i+1}  } 
+
\sum_{k=1}^{i-1}  \cd_{k,i+1} \cc_{ki} q^{ p_{i}\n_{[1,k-1],i} - p_{i+1}\n_{[1,k-1],i+1} } 
\nonumber \\
 & \qquad \text{for} \quad i+1 \in I,
\nonumber \\[6pt]
e_{i1}&=\theta(i \in I) \cd_{1i}  q^{ - p_{1}\n_{1,[2,i-1]}} 
 \quad \text{for} \quad  i \in{ \mathfrak I}\setminus \{1\},
 \nonumber
\end{align}
where the limit of $\tilde{e}_{ij}$ is denoted  again as $e_{ij}$. 
We remark that the relation $\overline{e}_{ii}=-e_{ii}$ holds only for $i \in I$ 
after the limit, and $q^{p_{i}\overline{e}_{ii}}=0$ for $i \in \overline{I}$ 
  means  that the contraction $\overline{L}_{ii}=0$ for $i \in \overline{I}$ 
occurs in the limit ($\overline{e}_{ii}$ for $i \in \overline{I}$ diverges and does not exist). 
Moreover, 
taking note on the relation \eqref{eij} in the limit, one can show 
\begin{align}
e_{ij} &= 0 \quad \text{for} \quad i,j \in \overline{I}, \quad i>j.
\label{real44e=0}
\end{align}
The other elements $e_{ij}$ can be obtained in two steps: 
$\{e_{ij}\}_{i<j}$ follow from $\{e_{i,i+1}\}_{i=1}^{M+N-1}$ 
 based on \eqref{a2} recursively; $\{e_{ic}\}_{i \in I,2\le c \le i-1}$ 
follow from $\{e_{i1}\}_{i \in I}$, $\{e_{ii}\}_{i \in {\mathfrak I}}$ and $\{e_{1c}\}_{c \ge 2}$ via \eqref{a9}. Then one can calculate:
\begin{align}
e_{ij} &=[e_{i,i+1},[e_{i+1,i+2},\dots ,[e_{j-2,j-1},e_{j-1,j}]_{q^{-p_{j-1}}} 
\dots ]_{q^{-p_{i+2}}}]_{q^{-p_{i+1}}}
\quad \text{for} \quad  i<j, 
 \label{other1}
\\[6pt]
e_{ic}&=q^{-p_{1}e_{11}+p_{c}e_{cc}}[e_{i1},e_{1c}] 
 \nonumber \\[3pt]
&=q^{-p_{1}e_{11}+p_{c}e_{cc}}[e_{i1},
[e_{12},[e_{23},\dots ,[e_{c-2,c-1},e_{c-1,c}]_{q^{-p_{c-1}}} 
\dots ]_{q^{-p_{3}}}]_{q^{-p_{2}}}] 
\nonumber \\[3pt]
& \qquad \qquad \text{for} \quad i \in I, \quad 2\le c \le i-1. 
\label{other2}
\end{align} 
We also remark that $\{e_{i1}\}_{i \in I,2 \le i<M+N}$ follow from $e_{M+N,1}$ based on \eqref{a10}:
\begin{align}
e_{i1}&=[e_{i,M+N},e_{M+N,1}]q^{-p_{M+N}e_{M+N,M+N}+p_{i}e_{ii}} 
\nonumber \\[3pt]
&=[[e_{i,i+1},[e_{i+1,i+2},\dots ,[e_{M+N-2,M+N-1},e_{M+N-1,M+N}]_{q^{-p_{M+N-1}}} 
\dots ]_{q^{-p_{i+2}}}]_{q^{-p_{i+1}}},e_{M+N,1}]
\nonumber \\[3pt]
& \qquad   \times 
q^{-p_{M+N}e_{M+N,M+N}+p_{i}e_{ii}} 
 \qquad \text{for} \quad i \in I, \quad 2\le i < M+N. 
\label{other3}
\end{align}
Thus we need only $\{e_{i,i+1}\}_{1 \le i \le M+N-1}$, 
$\{e_{ii}\}_{1 \le i \le M+N}$ and $e_{M+N,1}$ 
to calculate all the matrix elements of the L-operator in \eqref{rll-ev1} 
with \eqref{maprll1}-\eqref{maprll3}, \eqref{red300e} and \eqref{red30e}.  
The expression already \eqref{real44lim} realizes the contracted algebra $U_{q}(gl(M|N);I)$.  
We can simplify this more by removing the unnecessary parts. 
All the elements of the q-oscillator algebra super-commute among themselves 
if they have different indices. Thus 
the action of the terms containing any of the operators in $\{\cc_{ij} \}_{i,j \in \overline{I}}$ 
and $\{\cd_{ij} \}_{i,j \in \overline{I}}$
vanishes on the vacuum vector. 
Then we drop these terms from \eqref{real44lim} by formally setting
\footnote{The action of $\n_{ij}$ also vanishes if there is no action of $\cd_{ij}$.}  
\begin{align}
\cc_{ij} \mapsto 0,  \quad \cd_{ij} \mapsto 0,  \quad \n_{ij} \mapsto 0
 \quad \text{for} \quad i,j \in \overline{I},
  \label{redosc1}
\end{align}
 to get
\begin{align}
e_{ii} & = - \n_{i,I}
, \quad 
q^{p_{i}\overline{e}_{ii}}=0
 \quad \text{for} \quad   i \in \overline{I}, 
 \nonumber
\\[6pt]
e_{ii} & =\lambda_{i} +\n_{[1,i-1],i} - \n_{i,[i+1,M+N]}
, \quad 
\overline{e}_{ii}=-e_{ii}
 \quad \text{for} \quad   i \in I, 
 \nonumber
\\[6pt]
e_{i,i+1} &=  
 -p_{i}
\sum_{k \in I}  p_{k}
\cc_{ik} \cd_{i+1,k} q^{
- p_{i}\n_{i,[k,M+N]} +p_{i+1}\n_{i+1,[k,M+N]} +p_{i}+p_{i+1} }
\nonumber
\\
 & \qquad \text{for} \quad i,i+1 \in \overline{I},
 \nonumber
\\[6pt]
e_{i,i+1} &=  
p_{i}(q-q^{-1})^{-1}\cc_{i,i+1} 
q^{- p_{i+1} \lambda_{i+1} - p_{i} \n_{i,[i+1,M+N]} +
 p_{i+1} \n_{i+1,[i+2,M+N]}  +p_{i}
}
\nonumber
 \\
& \quad 
 -p_{i}
\sum_{k=i+2}^{M+N}  p_{k}
\cc_{ik} \cd_{i+1,k} q^{-p_{i+1}\lambda_{i+1}
- p_{i}\n_{i,[k,M+N]} +p_{i+1}\n_{i+1,[k,M+N]} +p_{i}+p_{i+1} }
\nonumber
\\
 & \qquad \text{for} \quad i \in \overline{I},  \quad i+1 \in I, \quad (i=a), 
 \nonumber
\\[6pt]
e_{i,i+1} &=  
\sum_{k=1}^{i-1}
 \cd_{ki} \cc_{k,i+1}
 \nonumber
\\
& \quad \times 
q^{-p_{i}\lambda_{i}+ p_{i+1} \lambda_{i+1} -
p_{i} \n_{[k+1,i-1],i} + p_{i+1} \n_{[k+1,i],i+1}  +
 p_{i} \n_{i,[i+1,M+N]} - p_{i+1} \n_{i+1,[i+2,M+N]}  } 
 \label{real44lim2}
 \\
 &
 \quad 
+
p_{i}\cc_{i,i+1} 
\left[
p_{i}\lambda_{i}- p_{i+1} \lambda_{i+1}- p_{i} \n_{i,[i+1,M+N]} +
 p_{i+1} \n_{i+1,[i+2,M+N]}  +p_{i}
\right]_{q}
\nonumber
 \\
& \quad 
 -p_{i}
\sum_{k=i+2}^{M+N}  p_{k}
\cc_{ik} \cd_{i+1,k} q^{p_{i}\lambda_{i}-p_{i+1}\lambda_{i+1} 
- p_{i}\n_{i,[k,M+N]} +p_{i+1}\n_{i+1,[k,M+N]} +p_{i}+p_{i+1} },
\nonumber
\\
 & \qquad \text{for} \quad i,i+1 \in I,
 \nonumber
 \\[6pt]
e_{ij} &= 0 \quad \text{for} \quad i,j \in \overline{I}, \quad  i>j,
\nonumber
\\[6pt]
e_{i+1,i} &=  \cd_{i,i+1} q^{  - p_{i+1}\n_{[1,i-1],i+1}  } 
 \quad \text{for} \quad i \in \overline{I}, \quad i+1 \in I, \quad (i=a),
 \nonumber
\\[6pt]
e_{i+1,i} &=   
\cd_{i,i+1} q^{ p_{i}\n_{[1,i-1],i} - p_{i+1}\n_{[1,i-1],i+1}  } 
+
\sum_{k=1}^{i-1}  \cd_{k,i+1} \cc_{ki} q^{ p_{i}\n_{[1,k-1],i} - p_{i+1}\n_{[1,k-1],i+1} } 
\nonumber
\\
 & \qquad \text{for} \quad i,i+1 \in I,
 \nonumber
\\[6pt]
e_{i1}&= \cd_{1i}  q^{ - p_{1}\n_{1,[a+1,i-1]}} 
 \quad \text{for} \quad  
 i \in I,  \quad i>1.
\nonumber
\end{align}
This expression \eqref{real44lim2} (with \eqref{other1} and \eqref{other2}) 
realizes the contracted algebra $U_{q}(gl(M|N);I)$ and gives 
an evaluation representation of the q-super-Yangian satisfying  
\eqref{HWcont} through \eqref{maprll1}-\eqref{rll-ev1} (see \eqref{real44lim2L}). 
We remark that an additional condition $n_{jb}=0$ for $j,b \in \overline{I}$ 
should be imposed on \eqref{vecFock} in accordance with the reduction \eqref{redosc1}. 

Next we consider the case $\lambda_{i}=p_{i}\mu$ for $i \in I $. 
We start from \eqref{real44} with the reduction \eqref{rel44-res} and 
repeat the same procedure to derive \eqref{real44lim2}  from  \eqref{real44} 
[we use \eqref{renoGen22}, \eqref{qlim}, \eqref{cijreo} and \eqref{redosc1}], 
to get 
\begin{align}
e_{ii} & = - \n_{i,I}
 , \qquad 
q^{p_{i}\overline{e}_{ii}}=0
 \quad \text{for} \quad   i \in \overline{I}, 
\nonumber \\[6pt] 
e_{ii}  &=p_{i}\mu+\n_{\overline{I},i}
, \qquad 
\overline{e}_{ii}=-e_{ii}
 \quad \text{for} \quad   i \in I, 
\nonumber  \\[6pt]
e_{i,i+1} &=  
 -p_{i}
\sum_{k \in I}  p_{k}
\cc_{ik} \cd_{i+1,k} q^{
- p_{i}\n_{i,[k,M+N]} +p_{i+1}\n_{i+1,[k,M+N]} +p_{i}+p_{i+1} }
\nonumber  \\
 & \qquad \text{for} \quad i,i+1 \in \overline{I},
\nonumber  \\[6pt]
e_{i,i+1} &=  
p_{i}(q-q^{-1})^{-1}\cc_{i,i+1} 
q^{-\mu - p_{i} \n_{i,I}  +p_{i}
}
%
  \quad \text{for} \quad i \in \overline{I},  \quad i+1 \in I, \quad (i=a), 
\nonumber  \\[6pt]
e_{i,i+1} &=  
\sum_{k \in \overline{I}}
 \cd_{ki} \cc_{k,i+1}
q^{ -
p_{i} \n_{[k+1,a],i} + p_{i+1} \n_{[k+1,a],i+1}   } 
 \quad \text{for} \quad i,i+1 \in I,
\label{real44lim3}  \\[6pt]
e_{ij} &= 0 \quad \text{for} \quad i,j \in \overline{I}, \quad  i>j,
\nonumber  \\[6pt]
e_{i+1,i} &=  \cd_{i,i+1} q^{  - p_{i+1}\n_{[1,i-1],i+1}  } 
 \quad \text{for} \quad i \in \overline{I}, \quad i+1 \in I, \quad (i=a),
\nonumber  \\[6pt]
e_{i+1,i} &=   
\sum_{k \in \overline{I}}  
\cd_{k,i+1} \cc_{ki} q^{ p_{i}\n_{[1,k-1],i} - p_{i+1}\n_{[1,k-1],i+1} } 
 \quad \text{for} \quad i,i+1 \in I,
\nonumber  \\[6pt]
e_{i1}&= \cd_{1i}  q^{ - p_{1}\n_{1,[a+1,i-1]}} 
 \quad \text{for} \quad
  i \in I,  \quad 1 \in \overline{I},  \quad i>1.
\nonumber  
\end{align}
This expression \eqref{real44lim3} (with \eqref{other1} and \eqref{other2}) 
realizes the contracted algebra $U_{q}(gl(M|N);I)$ and gives 
an evaluation representation of the q-super-Yangian satisfying  
\eqref{HWcont} with $\lambda_{i}=p_{i}\mu$ for $i \in I $ through \eqref{maprll1}-\eqref{rll-ev1} 
(see \eqref{real44lim3L}). 
We remark that 
this is equivalent to  \eqref{real44lim2} with the reduction \eqref{rel44-res}.  
We find that \eqref{real44lim3} for $\mu=0$ gives 
q-oscillator representations for Baxter Q-operators. 
Substituting these into \eqref{eva}, we obtain q-oscillator 
realization of a contracted algebra for $U_{q}(\hat{gl}(M|N))$:
\begin{align}
\begin{split}
k_{i} & = - \n_{i,I}
 \quad \text{for} \quad   i \in \overline{I}, 
\qquad 
k_{i}  =p_{i}\mu+\n_{\overline{I},i} 
 \quad \text{for} \quad   i \in I, 
\\[6pt]
e_{i} &=  
 -p_{i}
\sum_{k \in I}  p_{k}
\cc_{ik} \cd_{i+1,k} q^{
- p_{i}\n_{i,[k,M+N]} +p_{i+1}\n_{i+1,[k,M+N]} +p_{i}+p_{i+1} }
\\
 & \qquad \text{for} \quad i,i+1 \in \overline{I},
\\[6pt]
e_{i} &=  
p_{i}(q-q^{-1})^{-1}\cc_{i,i+1} 
q^{-\mu - p_{i} \n_{i,I}  +p_{i}
}
%
  \quad \text{for} \quad i \in \overline{I},  \quad i+1 \in I, \quad (i=a), 
\\[6pt]
e_{i} &=  
\sum_{k \in \overline{I}}
 \cd_{ki} \cc_{k,i+1}
q^{ -
p_{i} \n_{[k+1,a],i} + p_{i+1} \n_{[k+1,a],i+1}   } 
 \quad \text{for} \quad i,i+1 \in I,
\\[6pt]
e_{M+N}&= x \cd_{1,M+N}  q^{p_{1}-\mu + p_{1}\n_{1,M+N}-p_{M+N}\n_{\overline{I},M+N}} ,
\end{split}
\label{contaff441}
\end{align}
and 
\begin{align}
\begin{split}
f_{i} &= 0 \quad \text{for} \quad i+1 \in \overline{I},
\\[6pt]
f_{i} &= p_{i} \cd_{i,i+1} q^{  - p_{i+1}\n_{[1,i-1],i+1}  } 
 \quad \text{for} \quad i \in \overline{I}, \quad i+1 \in I, \quad (i=a),
\\[6pt]
f_{i} &=   p_{i}
\sum_{k \in \overline{I}}  
\cd_{k,i+1} \cc_{ki} q^{ p_{i}\n_{[1,k-1],i} - p_{i+1}\n_{[1,k-1],i+1} } 
 \quad \text{for} \quad i,i+1 \in I,
\\[6pt]
f_{M+N} &=p_{M+N}x^{-1}q^{p_{M+N}k_{M+N}}[e_{1},[e_{2},\dots ,
[e_{M+N-2},e_{M+N-1}]_{q^{-p_{M+N-1}}}\dots ]_{q^{-p_{3}}}]_{q^{-p_{2}}}
q^{p_{1}k_{1}} ,
\end{split}
\label{contaff442}
\end{align}
where $1 \in \overline{I}, M+N \in I$ is assumed.
In fact, these satisfy the following contracted commutation relations (cf. eq.(3.85) in \cite{Tsuboi12})
 instead of  the relations \eqref{efk-1}. 
\begin{align}
[e_{i},f_{j}]=\delta_{ij}\frac{\theta(i+1 \in I)q^{h_{i}}-\theta(i \in I)q^{-h_{i}}}{q-q^{-1}},
\qquad
h_{i}=p_{i}k_{i}-p_{i+1}k_{i+1}, 
\qquad i,j \in {\mathfrak I},
\label{contractedef}
\end{align}
where $M+N+1 \equiv 1$. 
The other relations \eqref{efk-0} and \eqref{com-ee} (and Serre type relations) remain valid. 
In addition, simplified Serre type relations may also hold 
(see \cite{BHK02} for ${\mathcal B}_{+}$ of $U_{q}(\hat{sl}(3))$, and \cite{Tsuboi12} 
for $U_{q}(\hat{gl}(M|N))$ case). 
In particular, 
\eqref{contaff441} realizes the Borel subalgebra $\mathcal{B}_{+}$ of 
the quantum affine superalgebra $U_{q}(\hat{gl}(M|N))$. On the Fock space, this gives 
q-oscillator representations for Baxter Q-operators. 
In fact,  special cases of  \eqref{contaff441} (in different conventions)  can be seen, 
for example in:
\cite{BLZ97}: for $I=\{2\}$, $M=2,N=\mu=0$; 
\cite{BHK02}: for $I=\{2,3\},\{3\}$, $M=3,N=\mu=0$; 
\cite{Kojima08} 
 for $I=\{2,3,\dots,M\}$,  $\{M\}$ and $N=\mu=0$; 
 \cite{BT08}: for $I=\{2,3\},\{3\}$, $M=2,N=1,\mu=0$;
\cite{Tsuboi12} for $I=\{2,3,\dots,M+N\}$,  $\{M+N\}$ and $\mu=0$, $N,M>0$. 
In addition,  the result of \cite{Kojima08} 
(\eqref{contaff441} for $I=\{M\}$ and $N=\mu=0$)
 was rederived
 \footnote{
 Set $\cc_{j,M}=(q-q^{-1})q^{{\mathcal H}_{j}+1}\varepsilon^{*}_{j}$, 
 $\cd_{j,M}=\varepsilon_{j}$, 
 $\n_{j,M}={\mathcal H}_{j}$ for $1 \le j \le M-1$, and apply 
 the automorphism  of ${\mathcal B}_{+}$: 
 $e_{1} \mapsto tq^{-\frac{1}{2}}e_{1}$,  
 $e_{j} \mapsto q^{-\frac{1}{2}}e_{j}$ 
 for $2 \le j \le M-2$,  $e_{M-1} \mapsto q^{-1}e_{M-1}$, 
 $e_{M} \mapsto x^{-1}q^{-1}e_{M}$, 
 $h_{j} \mapsto h_{j}$ for $1\le j \le M$ 
 to  \eqref{contaff441} for $I=\{M\}$ and $N=\mu=0$
 (we use the Cartan elements $h_{i}$ in \eqref{contractedef}; 
 $\varepsilon^{*}_{j},\varepsilon_{j},{\mathcal H}_{j},t$ 
 are symbols in \cite{Kojima08}). 
 Then one obtains eq. (2.2) in \cite{Kojima08} after 
 the transformation  $q \to q^{-1}$
 (Note that $N$ in \cite{Kojima08} corresponds to $M$, and 
 the central element of the q-oscillator algebra is fixed in this paper, while 
 it is free in \cite{Kojima08}).
 Next, 
 apply the automorphism of ${\mathcal B}_{+}$: $e_{i} \mapsto q^{-1}e_{i}$, 
 $k_{i} \mapsto k_{i}$ for $i \in {\mathfrak I}$
 to 
 \eqref{contaff441} for $I=\{M\}$ and $N=\mu=0$ 
 (we use the Cartan elements $h_{i}$ in \eqref{contractedef}). 
 Then apply the transformation $q \mapsto q^{-1}$ 
 and set $x \to 1$. One will find the homomorphism $\rho$ in page 15, section 8 in \cite{NR16}.}
  in \cite{NR16} by taking asymptotic limit of a  
Verma module of  $\mathcal{B}_{+}$ and factoring out invariant subspaces. 
Moreover, 
the same type of representations of $\mathcal{B}_{+}$ 
can be derived systematically as asymptotic limit of Kirillov-Reshetikhin modules 
(see \cite{HJ11} for $N=\mu=0$ case, and \cite{Zhang15,Zhang14} for $M,N>0$, $\mu=0$ case). 

It is easy to calculate all the generators of $U_{q}(gl(M|N;I))$ explicitly for $a=1$ and $M+N-1$ 
from \eqref{real44lim3}.
\\ 
The case $a=1$, $I=\{2,3,\dots,M+N\}$:
\begin{align}
e_{11}&=-\n_{1,I}, 
\qquad e_{ii}=p_{i}\mu+\n_{1i} \quad \text{for} \quad i \in I,
\nonumber \\[6pt]
e_{1j}&=p_{1}(q-q^{-1})^{-1}\cc_{1j}q^{-\mu-p_{1}\n_{1,[j,M+N]}+p_{1}} 
\quad \text{for} \quad j \in I,
\nonumber \\[6pt]
e_{ij}&=\cd_{1i}\cc_{1j}q^{p_{1}\n_{1,[i+1,j-1]}} 
\quad \text{for} \quad 2 \le i<j \le M+N,
\label{eija=1lim}
\\[6pt]
e_{i1}&=\cd_{1i}q^{-p_{1}\n_{1,[2,i-1]}} 
\quad \text{for} \quad i \in I,
\nonumber \\[6pt]
e_{ij}&=\cd_{1i}\cc_{1j}q^{-p_{1}\n_{1,[j+1,i-1]}} 
\quad \text{for} \quad 2 \le j<i \le M+N.
\nonumber 
\end{align}
The case $a=M+N-1$,  $I=\{M+N\}$:
\begin{align}
\begin{split}
e_{ii}&=-\n_{i,M+N} \quad \text{for} \quad i\in \overline{I}, 
\qquad e_{M+N,M+N}=p_{M+N}\mu+\n_{\overline{I},M+N},
\\[6pt]
e_{ij}&=-p_{i}p_{M+N}\cc_{i,M+N}\cd_{j,M+N}
q^{-p_{i}\n_{i,M+N}+p_{j}\n_{j,M+N}-p_{M+N}\n_{[i+1,j-1],M+N}+p_{i}+p_{j}} 
  \\
&  \qquad \text{for} \quad 1 \le i<j < M+N,
\\[6pt]
e_{i,M+N}&=p_{i}(q-q^{-1})^{-1}\cc_{i,M+N}q^{-\mu-p_{i}\n_{i,M+N}-p_{M+N}n_{[i+1,M+N-1],M+N}+p_{i}} 
\quad \text{for} \quad i \in \overline{I},
\\[6pt]
e_{M+N,j}&=\cd_{j,M+N}q^{-p_{M+N}\n_{[1,j-1],M+N}} 
\quad \text{for} \quad j \in \overline{I},
\\[6pt]
e_{ij}&=0
\quad \text{for} \quad 1 \le j<i < M+N.
\end{split}
\label{eija=K-1lim}
\end{align}
One can also derive 
\eqref{eija=1lim} directly from \eqref{eija=1}
 in the limit \eqref{qlim} with \eqref{renoGen22} 
 and \eqref{cijreo}. 
Substituting \eqref{eija=1lim} or \eqref{eija=K-1lim} 
into the expression $\Lf(x)$ in \eqref{rll-ev1} with \eqref{maprll1}-\eqref{maprll3}, \eqref{red300e} 
and $\mu=0$, 
we obtain L-operators for Q-operators (see Appendix D for these types of L-operators in different conventions). 
\section{Rational case}
In this section, we will discuss the rational case. 
We will present a factorization formula of the L-operator for $Y(gl(M|N))$, 
which is a generalization of  the results in \cite{D05,DM06,DM10}. 
By taking limits of the L-operator, we 
 recover the rational L-operators for Q-operators proposed in \cite{FLMS10,BFLMS10}.

In the rational limit $q \to 1$, \eqref{qosc} reduces to
\begin{align}
\begin{split}
& [\cc_{ia}, \cd_{jb}]  =\delta_{ab} \delta_{ij}  , 
\\[6pt]
&  
[\n_{ia}, \cc_{jb}]=-\delta_{ij}\delta_{ab} \cc_{jb}, \quad 
 [\n_{ia}, \cd_{jb}]=\delta_{ij}\delta_{ab} \cd_{jb}, \quad
[\n_{ia}, \n_{jb}]=[\cc_{ia}, \cc_{jb}]=[\cd_{ia}, \cd_{jb}]=0.
\end{split}
\label{osc}
\end{align}
where the Cartan elements $\n_{ia}$ are realized as  
 $ \cc_{ia}\cd_{ia}=1+ p_{i}p_{a}\n_{ia}$, 
$ \cd_{ia}\cc_{ia}=\n_{ia}$. 
Then the rational limits of  \eqref{real44} and \eqref{enj-44} with \eqref{eij} 
 are  given by 
\begin{align}
\begin{split}
e_{ii} & =\lambda_{i} +\n_{[1,i-1],i} - \n_{i,[i+1,M+N]} 
\quad  \text{for} \quad  i \in \Ik, 
\\[6pt]
e_{i,i+1} &= 
\sum_{k=1}^{i-1} \cd_{ki} \cc_{k,i+1} 
\\
 &
 \quad 
+
p_{i}\cc_{i,i+1} 
(
p_{i}\lambda_{i}- p_{i+1} \lambda_{i+1}- p_{i} \n_{i,[i+1,M+N]} +
 p_{i+1} \n_{i+1,[i+2,M+N]}  +p_{i}
)
 \\
& \quad 
 -p_{i}
\sum_{k=i+2}^{M+N}  p_{k}
\cc_{ik} \cd_{i+1,k} \quad \text{for} \quad 
  i \in \Ik \setminus \{M+N\}, 
\\[6pt]
e_{ji} &= 
\cd_{ij} +\sum_{k=1}^{i-1}  \cd_{kj} \cc_{ki}
 \quad \text{for} \quad 
j>i, \quad i,j \in \Ik.
\end{split}
\label{real44r}
\end{align}
These expressions of generators can be written as 
 a factorized matrix form
\footnote{
We could not find this type of formula for $U_{q}(gl(M|N))$ for generic $(M,N)$ in literatures, 
and have obtained special cases of it at the moment. We leave this for future work.
}
%
$E=zDz^{-1}$,
where 
\begin{align}
\begin{split}
E&=\sum_{i,j \in {\mathfrak I}}p_{i}e_{ji}\otimes E_{ij},
\qquad
%
D=\sum_{i,j \in {\mathfrak I}}p_{i}(\delta_{ij}d_{i}+D_{ji})\otimes E_{ij},
\\[6pt]
z&=\sum_{i,j \in {\mathfrak I}}z_{ij} \otimes E_{ij}.
\qquad
z^{-1}=\sum_{i,j \in {\mathfrak I}}y_{ij} \otimes E_{ij}.
\end{split}
\label{EDz}
\end{align}
In components, it reads 
\begin{multline}
(-1)^{p(i)(p(j)+1)}e_{ij}=
\\=
\sum_{a,b\in {\mathfrak I}}
(-1)^{p(j)(p(a)+1)}z_{ja}
(-1)^{(p(a)+1)p(b)}(\delta_{ab}d_{a}+D_{ba})
(-1)^{(p(b)+1)p(i)}y_{bi} , 
\label{eijzDz}
\end{multline}
where each element is defined by 
\begin{align}
(-1)^{(p(b)+1)p(i)}y_{bi} &=-(-1)^{(p(b)+1)p(i)}z_{bi}
\nonumber \\
&  +
\sum_{k=2}^{b-i} (-1)^{k}
\sum_{b>a_{1}>a_{2}>\cdots > a_{k-1}>i}
(-1)^{(p(b)+1)p(a_{1})}z_{ba_{1}}
(-1)^{(p(a_{1})+1)p(a_{2})}z_{a_{1}a_{2}}
\nonumber  \\
&  \cdots
(-1)^{(p(a_{k-2})+1)p(a_{k-1})}z_{a_{k-2}a_{k-1}}
(-1)^{(p(a_{k-1})+1)p(i)}z_{a_{k-1}i}
 \quad \text{for} \quad b>i,
%
\nonumber \\[6pt]
 y_{ii}&=1, \qquad y_{bi}=0  \quad \text{for} \quad b<i,
\nonumber  \\[6pt]
z_{ij}&=p_{i}p_{j}\cc_{ji} \quad \text{for} \qquad i>j, \qquad z_{ii}=1, 
\qquad z_{ij}=0\quad \text{for} \quad i<j, 
\nonumber  \\[6pt]
D_{ij}&=\cd_{ji}+p_{i}\sum_{k=i+1}^{M+N}p_{k}\cc_{ik}\cd_{jk}, 
\quad \text{for}\quad  i>j, \qquad D_{ij}=0 \quad \text{for}\quad  i \le j,
\nonumber  \\[6pt]
d_{a}&=\lambda_{a}-\sum_{k=1}^{a-1}p_{k}p_{a} .
\label{ybi}
\end{align}
Due to the graded tensor product, the condition $z z^{-1}= z^{-1}z=1\otimes 1$ produces an 
extra sign factor 
\begin{align}
\sum_{k\in {\mathfrak I}}(-1)^{(p(i)+p(k))(p(k)+p(j))}z_{ik}y_{kj}=
\sum_{k\in {\mathfrak I}}(-1)^{(p(i)+p(k))(p(k)+p(j))}y_{ik}z_{kj}=
\delta_{ij}. 
\label{z-y}
\end{align} 
In short, 
the matrices 
\footnote{instead of 
$(z_{ij})_{1\le i,j \le M+N}$
and $(y_{ij})_{1\le i,j \le M+N}$} $((-1)^{(p(i)+1)p(j)}z_{ij})_{1\le i,j \le M+N}$ 
and $((-1)^{(p(i)+1)p(j)}y_{ij})_{1\le i,j \le M+N}$ have the normal matrix product. 
We remark that the elements $D_{ij}$ for $i>j$ satisfy the relations 
$[D_{ij},D_{kl}]=-\delta_{jk}D_{il}+(-1)^{(p(i)+p(j))(p(k)+p(i))}\delta_{li}D_{kj}$ 
for $i>j$ and $k>l$, and thus  $-D_{ij}$ for $i>j$ obey the relations for $gl(M|N)$. 
We also have 
$[z_{ij},D_{kl}]=p_{i}p_{j}\delta_{ik}\delta_{jl}+(-1)^{(p(i)+p(j))(p(i)+p(k)+1)}\delta_{jl}\theta(i>k)z_{ik}$ 
for $i>j$ and $k>l$.  
Based on these relations, one can check that 
\eqref{eijzDz} satisfies the relations for $gl(M|N)$. 
\begin{align}
[e_{ij},e_{kl}]=\delta_{jk}e_{il}-
  (-1)^{(p(i)+p(j))(p(k)+p(l))} \delta_{li}  e_{kj} 
  \quad \text{for} \quad 
  i,j,k,l \in {\mathfrak I}.
   \label{glM}
\end{align}
The above types of factorization formulas are known in \cite{D05} for $sl(2|1)$ 
and in \cite{DM06} for $sl(N)$. See also section 5.3 in \cite{DM10} 
for a review on $gl(N)$ case. 
We also remark that 
 the unitary representations of the non-compact real forms of $sl(M|N)$ 
 are studied in \cite{GV17} based on another oscillator realization of the algebra. 
By using the relation \eqref{z-y}, we can show 
\begin{multline}
\sum_{\beta=i}^{M+N}(-1)^{p(\beta)(p(\alpha)+1)}D_{\beta \alpha}
(-1)^{(p(\beta)+1)p(i)}y_{\beta i}=
\\[6pt]
=
\begin{cases}
(-1)^{(p(\alpha)+1)p(i)}\cd_{\alpha i} 
& \text{for} \quad \alpha <i ,
\\[6pt]
-\n_{i,[i+1,M+N]} 
  & \text{for} \quad \alpha =i ,
\\[6pt]
(-1)^{p(i+1)p(i)}\bigl(-\sum_{k=i+2}^{M+N}(-1)^{p(k)}\cc_{ik}\cd_{i+1,k}
&+(-1)^{p(i+1)}\cc_{i,i+1}\n_{i+1,[i+2,M+N]} \bigr)
\\
&  \text{for} \quad \alpha =i+1 .
\end{cases}
\label{Dy=c}
\end{multline}
Then, 
applying  \eqref{Dy=c} to \eqref{eijzDz}, we get 
\eqref{real44r}. 
Let us consider 
the rational limits of the R- and L-operators (defined in  \eqref{PS-R} and \eqref{rll-ev1}):
\begin{align}
 R(u)&=(q-q^{-1})^{-1}\lim_{q \to 1}\Rf (q^{-2u})=
 u(1\otimes 1)+\sum_{i,j\in {\mathfrak I}}p_{i}E_{ji}\otimes E_{ij},
 \\[6pt]
 L(u)&=(q-q^{-1})^{-1}\lim_{q \to 1}\Lf (q^{2u})=
 u(1\otimes 1)+\sum_{i,j\in {\mathfrak I}}p_{i}e_{ji}\otimes E_{ij},
 \label{Lop-ra}
\end{align}
where $u \in {\mathbb C}$. 
These satisfy the following Yang-Baxter relation, which is the rational limit of \eqref{YB-L}.
\begin{align}
& R^{23}(u-v)L^{13}(v)L^{12}(u)=
L^{12}(u)L^{13}(v)R^{23}(u-v), 
\qquad 
u,v \in {\mathbb C}.
\label{YB-Lra}
\end{align}
Using \eqref{EDz}, we obtain a factorization formula for the L-operator \eqref{Lop-ra}:
\begin{align}
L(u)=z(u(1\otimes 1) +D)z^{-1}
\end{align}
This is a generalization of the factorization formulas \cite{D05,DM06,DM10} 
to the case $Y(gl(M|N))$. 
Let us take a subset $I=\{a+1,a+2,\dots, M+N \}$ of ${\mathfrak I}$ and 
it complement set $\overline{I}={\mathfrak I} \setminus I $. 
Then we consider \eqref{real44r} or \eqref{EDz} for the case 
$\lambda_{i}=p_{i}m$ for $i \in \overline{I}$, 
and rewrite them  
in the following form 
(use the relations \eqref{z-y} and \eqref{Dy=c}). 
\begin{align}
e_{ij}&=p_{i}m\delta_{ij}+o(m), \quad i,j \in \overline{I},
\nonumber 
\\[6pt]
e_{ij} &= 
p_{i} m \cc_{ij}+
  m \sum_{k \in \overline{I}, k > i} p_{k} y_{ki} \cc_{kj} +o(m)
\quad 
 \text{for} \quad i<j, \quad i \in \overline{I}, \quad  j \in I, 
 \nonumber 
\\[6pt]
e_{ij} &=  
\cd_{ji} 
+
\sum_{k=1}^{j-1}  \cd_{ki} \cc_{kj} 
\quad \text{for} \quad 
j<i, \quad  j \in \overline{I},  \quad i \in I,  
\label{real44rm} 
\\[6pt]
e_{ij} &= 
e^{I}_{ij}
 +
\sum_{k \in \overline{I}}  \cd_{ki} \cc_{kj} 
\quad  
\text{for} \quad i,j \in I ,
\nonumber 
\end{align}
where $y_{ki}$ is a function of $\{ \cc_{\alpha \beta}\}_{i \le  \alpha < \beta \le k}$ 
and is linear 
 with respect to each 
$\cc_{\alpha \beta}$ (see \eqref{ybi}); $o(m)$ denotes the terms which do not depend on $m$; 
$\{e^{I}_{ij}\}_{i,j \in I}$ are the terms in $e_{ij}$ whose indices of the 
oscillator algebra are restricted to the set $I$. 
Note that  $\{e^{I}_{ij}\}_{i,j \in I}$  realizes a subalgebra of  $gl(M|N)$, which 
we denote
\footnote{$gl(I)=gl(\tilde{M}| \tilde{N})$, where 
$\tilde{M}=\mathrm{Card}\{j\in I| p(j)=0 \}$, 
$\tilde{N}=\mathrm{Card}\{j\in I| p(j)=1 \}$.} as $gl(I)$, and 
on the Fock space,  gives a highest weight 
representation with the highest weight $(\lambda_{a+1},\dots,\lambda_{M+N})$.
We renormalize the oscillator realization  \eqref{real44rm} as
\begin{align}
\tilde{e}_{ij}=
\left(m^{-1}\theta(i \in \overline{I}) + \theta(i \in I) \right) e_{ij}. 
\end{align}
Then we find that 
the limit $\lim_{m \to \infty}\tilde{e}_{ij}$, which is denoted again as $e_{ij}$, 
 satisfies the following contracted commutation relations:
\begin{align}
[e_{ij},e_{kl}]=\delta_{jk}\theta(j,k \in I)e_{il}-
  (-1)^{(p(i)+p(j))(p(k)+p(l))} \delta_{li} \theta(l,i\in I) e_{kj} .
   \label{cont-rat}
\end{align}
Explicitly, we obtain
\begin{align}
e_{ij}&=p_{i}\delta_{ij} \quad \text{for} \quad i,j \in \overline{I}, 
\nonumber  \\[6pt]
e_{ij}&=p_{i}  \cc_{ij}+
   \sum_{k \in \overline{I}, k > i}p_{k} y_{ki} \cc_{kj} 
 \quad \text{for} \quad i \in \overline{I},  \quad j \in I,
\nonumber  \\[6pt]
e_{ij} &=  
\cd_{ji} 
+
\sum_{k=1}^{j-1}  \cd_{ki} \cc_{kj} 
 \quad \text{for} 
\quad 
 i \in I, \quad j \in \overline{I},  
\label{real44rm-lim}  \\[6pt]
e_{ij} &= 
e^{I}_{ij}
+
\sum_{k \in \overline{I}}  \cd_{ki} \cc_{kj} 
\quad  
\text{for} \quad 
i,j \in I.
\nonumber  
\end{align}
Note that \eqref{real44rm-lim} does not depend on the generators $\{\cd_{ij}\}_{i,j\in \overline{I}}$.
Then, 
without breaking the relations \eqref{cont-rat}, we can forget about them and 
formally set 
their counterparts to zero: 
\begin{align}
\cc_{ij} \mapsto 0 \quad \text{for} \quad i,j \in \overline{I}. 
\label{cc=0}
\end{align}
Then \eqref{real44rm-lim}  reduces to 
\begin{align}
\begin{split}
e_{ij}&=p_{i}\delta_{ij} \quad \text{for} \quad i,j \in \overline{I}, 
\\[6pt]
e_{ij}&=p_{i}  \cc_{ij} 
 \quad \text{for} \quad i \in \overline{I},  \quad j \in I,
\\[6pt]
e_{ij} &=  
\cd_{ji} 
 \quad \text{for} 
\quad 
 i \in I, \quad j \in \overline{I},  
\\[6pt]
e_{ij} &= 
e^{I}_{ij}
+
\sum_{k \in \overline{I}}  \cd_{ki} \cc_{kj} 
\quad  
\text{for} \quad 
i,j \in I. 
\end{split}
\label{real44rm-lim-re} 
\end{align}
In case the vacuum vector $|0 \rangle $ is defined by $\cc_{ij}|0 \rangle=0$ (for any $i<j$), 
the parts depending on 
$\{\cc_{ij}\}_{i,j \in \overline{I}; i<j}$ vanish on the Fock space 
since $\{\cc_{ij}\}_{i,j \in \overline{I}; i<j}$ super-commute with any 
 elements in \eqref{real44rm-lim}. 
This justifies the reduction \eqref{cc=0}. 
Moreover, $\{e^{I}_{ij}\}_{i,j \in I; i<j}$ in \eqref{real44rm-lim-re} 
 super-commute with all the generators $\{\cc_{ij},\cd_{ij}| (i,j) \notin I \times I \}$ 
 of the oscillator the algebra. Then \eqref{real44rm-lim-re} satisfies 
 the relations \eqref{cont-rat} even if $\{e^{I}_{ij}\}_{i,j \in I}$ 
are replaced by the generic generators of $gl(I)$ 
($\{ e_{ij}\}$ should be interpreted as elements in the direct sum of $gl(I)$ and 
 the oscillator algebra). 

Let us introduce a diagonal matrix 
$g_{m}=\sum_{i =1}^{N}\left(m^{-1}\theta(i \in \overline{I}) + \theta(i \in I) \right) E_{ii}$. 
Then we take the limit of a renormalized version of L-operator \eqref{Lop-ra} 
with $\lambda_{i}=p_{i}m$ for $i \in \overline{I}$
(cf.\ \cite{BLMS10} for $(M,N)=(2,0)$ case):
\begin{align}
{\mathsf L}_{I}(u)=
\lim_{m\to \infty}L(u)(1 \otimes g_{m})|_{\eqref{cc=0}}=
u\sum_{i \in I}1 \otimes E_{ii}+
\sum_{i,j=1}^{M+N}p_{j}e_{ij} \otimes E_{ji},
\label{LopQra}
\end{align}
where $e_{ij}$ are defined in \eqref{real44rm-lim-re}. 
This satisfies the limit of the Yang-Baxter relation \eqref{YB-Lra}: 
\begin{align}
& R^{23}(u-v){\mathsf L}_{I}^{13}(v){\mathsf L}_{I}^{12}(u)=
{\mathsf L}_{I}^{12}(u){\mathsf L}_{I}^{13}(v)R^{23}(u-v)
\label{YB-Lra-lim}
\end{align}
since the relation $R(u)(g_{m} \otimes g_{m})=(g_{m} \otimes g_{m})R(u)$ 
holds for any $m,u \in {\mathbb C}$, 
and the reduction \eqref{cc=0} keeps the relation \eqref{cont-rat} unchanged. 
The L-operator \eqref{LopQra} coincides
\footnote{
Make the shift 
$e^{I}_{ij}\mapsto e^{I}_{ij} -\sum_{k\in \overline{I}}(-1)^{p(k)+p(i)}\delta_{ij}/2$ 
(namely, $\lambda_{i}\mapsto \lambda_{i} -\sum_{k\in \overline{I}}(-1)^{p(k)+p(i)}/2$ 
for $i \in I$ in $e^{I}_{ij}$), 
 regard $\{e^{I}_{ij}\}_{i,j \in I}$ as the generic generators of $gl(I)$, 
apply the automorphism $e^{I}_{ij}\mapsto -(-1)^{p(j)+p(i)p(j)}e^{I}_{ji}$
 of $gl(I)$, and the automorphism 
$\cc_{ij} \mapsto (-1)^{p(j)+p(i)p(j)}\cd_{ij},\cd_{ij} \mapsto -(-1)^{p(i)+p(i)p(j)}\cc_{ij}$ 
of the oscillator algebra 
 to \eqref{LopQra}.
} 
with the L-operator proposed in \cite{FLMS10} (and for $Y(gl(M))$, see \cite{BFLMS10}) 
if  $\{e^{I}_{ij}\}_{i,j \in I}$ are interpreted as the generic generators of $gl(I)$. 
It defines an evaluation representation of a degenerated Yangian. 
In particular, when the $gl(I)$ part is trivial, namely $e^{I}_{ij}=0$, 
the L-operator  \eqref{LopQra} gives the L-operators for Q-operators \cite{FLMS10}. 
The requirement $e^{I}_{ij}=0$ (in addition to \eqref{cc=0}) corresponds to formally  setting  
\begin{align}
\lambda_{k} \mapsto 0 \quad \text{for any } k \in I;
\quad \cc_{ij}\mapsto 0, \quad  \cd_{ij}\mapsto 0  
\quad \text{for} \quad (i,j) \notin \overline{I} \times I,
\end{align}
Instead, one may start from the rational limit of \eqref{real44} with 
the reductions \eqref{rel44-res} and $\mu=0$,  and consider the limit 
of the form \eqref{LopQra}.
\section{Concluding remarks}
In this paper, we have constructed q-oscillator realizations of 
the q-super-Yangian $Y_{q}(gl(M|N))$  for Baxter Q-operators  
based on the Heisenberg realization of $U_{q}(gl(M|N))$ \cite{Kimura96,AOS97} 
(and \cite{ANO94} for $N=0$ case). 
It is known that free field realization (Wakimoto construction) of $U_{q}(\hat{sl}(M|N))$ 
can be constructed based on this Heisenberg realization of $U_{q}(sl(M|N))$ 
(cf. \cite{Kojima17,AOS97}). 
It will be interesting to consider an opposite direction, namely to consider 
reductions and limits of free field realizations of the quantum affine superalgebras 
to get  q-oscillator realizations of 
the q-super-Yangians for Baxter Q-operators.  
This may give another
\footnote{other than Hernandez-Jimbo \cite{HJ11}} 
systematic approach  to the problem for the 
quantum affine superalgebras other than type A, 
where evaluation representations are not available. 

One of the unsolved problems related to our topics is fusion of the L-operators 
for Q-operators.  For the rational case \cite{BFLMS10,FLMS10} 
(see also \cite{D05,DM06,DM10} for a different approach),   
one can construct the L-operators for Verma modules from 
the L-operators for Q-operators by fusion procedures. 
As for the trigonometric case, we have fusion formulas \cite{KT14} on the level of 
 the universal L-operators
 \footnote{These are independent of the space (quantum space) on which 
 the operators act.} for Q-operators associate with $U_{q}(\hat{sl}(2))$. 
 However, 
  similar formulas for $U_{q}(\hat{gl}(M|N))$ (for general $M,N$) 
   have not been established yet. 

In \cite{FT17}, the Lax matrices for the Toda system were discussed in the context of 
`shifted Yangians' or `shifted quantum affine algebras'. 
Apparently, some of these Lax matrices have similar structures as L-operators 
for Q-operators. It will be desirable to clarify how our approach 
fits into their formulation.  
\section*{Acknowledgments} 
  The work of the author was 
 supported in part by  
 the Australian Research Council 
 (at  Department of Theoretical Physics, RSPE,  Australian National University in 2013, 
 and at School of Mathematics and Statistics, the University of Melbourne in 2014-2016), 
Fakult\"at f\"ur Mathematik und Naturwissenschaften, 
Bergische Universit\"at Wuppertal in 2016,  
CNRS (at 
Laboratoire de Math\'ematiques et Physique Th\'eorique CNRS/UMR 7350,
 F\'ed\'eration Denis Poisson FR2964,
Universit\'e de Tours in 2016), 
 the European Research Council 
[Programme ``Ideas'' ERC-2012-AdG 320769 AdS-CFT-solvable]
(at Laboratoire de physique th\'eorique, 
D\'epartement de physique 
de l'ENS, \'Ecole normale sup\'erieure in 2017-2018), and 
Osaka City University Advanced Mathematical Institute (MEXT Joint Usage/Research Center on  Mathematics and Theoretical Physics) in 2019. 
The author thanks the anonymous referee for useful comments. 
\section*{Appendix A: Relations for $U_{q}(gl(M|N))$ and $U_{q}(gl(M|N;I))$}
\label{relationsglmn}
\addcontentsline{toc}{section}{Appendix A}
\def\theequation{A\arabic{equation}}
One can rewrite the relations \eqref{rll0}-\eqref{rll4} in terms of $e_{ij}$ 
and $\eb_{ii}$ through
 \eqref{maprll1}-\eqref{maprll3} as follows.
\setcounter{equation}{0}
\begin{align}
& [q^{p_{a}e_{aa}},q^{p_{b}e_{bb}}]=[q^{p_{a}\eb_{aa}},q^{p_{b}\eb_{bb}}]
=[q^{p_{a}e_{aa}},q^{p_{b}\eb_{bb}}]=0,
\label{a00}
\\
& q^{p_{a}e_{aa}} q^{p_{a}\eb_{aa}}=q^{p_{a}\eb_{aa}} q^{p_{a}e_{aa}}=1,
\label{a0}
\\
& e_{ab}q^{p_{c}\eb_{cc}}q^{p_{c}e_{cc}}=
[e_{ac},e_{cb}]_{q^{p_{c}}} \quad \text{for} \quad a>c>b,
\label{a1}
\\
& e_{ab}=[e_{ac},e_{cb}]_{q^{-p_{c}}} \quad \text{for} \quad a<c<b,
\label{a2}
\\
& [e_{ab}, e_{ba}] =p_{a} 
 \frac{q^{p_{a} e_{aa}}q^{p_{b} \eb_{bb} } -q^{p_{a} \eb_{aa}}q^{p_{b}e_{bb} }  }{q-q^{-1}}
 \quad \text{for} \quad a <b,
\\
& [e_{dc}, e_{ba}] =(-1)^{p(a)p(b)+(p(a)+p(b))p(c)+1} (q-q^{-1}) e_{da} e_{bc} 
\quad \text{for} \quad b <d<a<c 
\nonumber \\
& \hspace{130pt} \text{or} \quad 
a<c<b<d ,
\\
& [e_{dc}, e_{ba}] =0
\quad \text{for} 
\quad d <c<b<a \quad \text{or}  
\quad d >c>b>a \quad \text{or} 
\quad d <b<a<c \quad \text{or} 
\nonumber 
\\
& 
\quad d >b>a>c \quad \text{or} 
\quad d <c \le a<b\quad \text{or} 
\quad c <d \le b<a\quad \text{or} 
\quad d <a < b<c \quad \text{or} 
\nonumber 
\\
& 
\quad c <b < a<d ,
\\
& [e_{dc}, e_{ba}] =(-1)^{p(a)p(b)+(p(a)+p(b))p(c)+1} (q-q^{-1}) 
 q^{p_{a} e_{aa} -p_{c} e_{cc} } e_{da} e_{bc} 
 \nonumber 
\\
& 
 \quad \text{for} \quad d <a<c<b,   
\\
& [e_{dc}, e_{ba}] =(-1)^{p(a)p(b)+(p(a)+p(b))p(c)} (q-q^{-1}) 
e_{da} e_{bc}  q^{p_{b} e_{bb} -p_{d} e_{dd} } 
 \nonumber 
\\
& 
 \quad \text{for} \quad a <d<b<c,     
\\
& [e_{ba}, e_{ac}] = e_{bc}  q^{p_{b} e_{bb}}q^{p_{a} \eb_{aa} } 
 \quad \text{for} \quad a <b<c,  
 \label{a8}
\\
& [e_{ba}, e_{ac}] =  q^{p_{a} e_{aa}-p_{c} e_{cc} } e_{bc} 
 \quad \text{for} \quad a <c<b, 
 \label{a9}   
\\
& [e_{db}, e_{ba}]  = e_{da}  q^{p_{b} e_{bb}-p_{d} e_{dd}}
 \quad \text{for} \quad a <d<b,  
 \label{a10}
\\
& [e_{db}, e_{ba}] = q^{p_{a} e_{aa}}q^{p_{b} \eb_{bb} } e_{da}  
 \quad \text{for} \quad d <a<b,
 \label{a11}  
\\
& [e_{da}, e_{ba}] _{q^{-p_{a}}}=0 
 \quad \text{for} \quad a <b<d
  \quad \text{or} \quad b <d<a,
\\
& [e_{bc}, e_{ba}] _{q^{p_{b}}}=0 
 \quad \text{for} \quad c <a<b
  \quad \text{or} \quad b <c<a,
\\
& [e_{ba},e_{ba}]=0, \label{a16}
\end{align}
where $a,b,c,d \in {\mathfrak I}$.
We use the convention used in Appendix A in \cite{Tsuboi12}. 
\eqref{a16} reduces to $(e_{ba})^{2}=0$ for $p_{a}p_{b}=-1$, 
and becomes trivial for $p_{a}p_{b}=1$.
The contracted algebra $U_{q}(gl(M|N;I))$ can be obtained by imposing the 
conditions  
\eqref{red300e} and \eqref{red30e}, and replacing \eqref{a0} with   
\begin{align}
q^{p_{c}\eb_{cc}}q^{p_{c}e_{cc}}=
q^{p_{c}e_{cc}}q^{p_{c}\eb_{cc}}=\theta(c\in I) .
\end{align}
Note that some of the relations 
become trivial ($0=0$) under the reductions. 
The original algebra 
$U_{q}(gl(M|N))$ corresponds to $U_{q}(gl(M|N;{\mathfrak I}))$, where 
the factor $q^{p_{c}\eb_{cc}}q^{p_{c}e_{cc}}$ in \eqref{a1} becomes 1. 
The contracted algebra $U_{q}(gl(M|N;I))$ for $\mathrm{Card} (I)=1,2$ 
 was proposed in \cite{Bp}  for $(M,N)=(3,0)$,  and in \cite{talks} 
 for $(M,N)=(2,1)$. 
\section*{Appendix B: general q-oscillator and Heisenberg realizations of $U_{q}(gl(M|N))$} 
\label{genos-app}
\addcontentsline{toc}{section}{Appendix B}
\def\theequation{B\arabic{equation}}
\setcounter{equation}{0}
In \cite{AOS97,Kimura96}, q-difference (Heisenberg) realization of $U_{q}(sl(M|N))$ 
was proposed (see, \cite{ANO94} for $U_{q}(sl(M))$ case). 
In this section, we transcribe their results for $U_{q}(gl(M|N))$ case 
in terms of the q-oscillator algebra. 
Let $\lambda_{i} \in {\mathbb C}$ ($i \in \Ik$). 
Then,  $U_{q}(gl(M|N))$ is realized by
\begin{align}
e_{ii} & =\lambda_{i} +\n_{[1,i-1],i} - \n_{i,[i+1,M+N]} 
\quad \text{for} \quad i \in \Ik , 
\nonumber \\[6pt]
e_{i,i+1} &= \cc_{i,i+1} q^{ -p_{i}\n_{[1,i-1],i} + p_{i+1}\n_{[1,i-1],i+1}  } 
+\sum_{k=1}^{i-1} \cd_{ki}\cc_{k,i+1} q^{ -p_{i}\n_{[1,k-1],i} + p_{i+1}\n_{[1,k-1],i+1} } ,
\nonumber  \\[6pt]
e_{i+1,i} &= 
\sum_{k=1}^{i-1} 
\cd_{k,i+1} \cc_{ki}  q^{p_{i}\lambda_{i}- p_{i+1} \lambda_{i+1} + 
p_{i} \n_{[k+1,i-1],i} - p_{i+1} \n_{[k+1,i],i+1}  -
 p_{i} \n_{i,[i+1,M+N]} + p_{i+1} \n_{i+1,[i+2,M+N]}  } 
\nonumber  \\
 &
 \quad 
+
p_{i}
\cd_{i,i+1} 
\left[
p_{i}\lambda_{i}- p_{i+1} \lambda_{i+1}- p_{i} \n_{i,[i+1,M+N]} + p_{i+1} \n_{i+1,[i+2,M+N]}  
\right]_{q}
\nonumber  \\
& \quad 
 -p_{i}
\sum_{k=i+2}^{M+N}
 p_{k}\cc_{i+1,k}\cd_{ik} q^{-p_{i}\lambda_{i}+p_{i+1}\lambda_{i+1} 
+ p_{i}\n_{i,[k,M+N]} -p_{i+1}\n_{i+1,[k,M+N]}  }
\nonumber \\ &
 \hspace{170pt}  \text{for} \quad i \in \Ik  \setminus \{M+N\}.
\label{real1}
\end{align}
The other generators can be obtained by the relations \eqref{eij}. 
In particular, the element $e_{1j}$ has quite a simple form
\footnote{The corresponding expression for $N=0$ case is written in \cite{ANO94} 
in terms of q-difference operators.}
\begin{align}
e_{1j}=
\cc_{1j} q^{ p_{1}\n_{1,[2,j-1]}  }, \qquad 2 \le j \le M+N.
 \label{e1j}
\end{align}
Let us consider reduction of the 
q-oscillator algebra in \eqref{real1}. 
Fix parameters $a \in \{0,1,\dots, M+N\}$ and $\mu \in {\mathbb C}$, and define a set by 
$I=\{a+1,a+2,\dots,M+N \}$. 
We find that 
\eqref{real1} still realizes $U_{q}(gl(M|N))$ 
even if we apply the following replacement:
\begin{align}
\cc_{ij} \mapsto 0 , 
\quad 
\cd_{ij} \mapsto 0,
\quad 
\n_{ij} \mapsto 0 ,
\quad 
\lambda_{i} \mapsto p_{i}\mu 
\quad \text{for} \quad i,j \in I. 
 \label{rel-res1}
\end{align}
This fact was remarked in \cite{ANO94} 
for $N=0$, $a=1$, $\mu=0$ case, where \eqref{real1} 
reduces to a q-analogue of the Holstein-Primakoff realization (cf. \cite{P98}). 

In this paper, we realize the algebra in terms of the q-oscillator superalgebras. 
One can rewrite these in terms of q-difference operators. 
Let us introduce variables $x_{ij}$ ($ 1 \le i<j \le M+N $) with the Grassmann parities $p_{i}p_{j}$ and define operators 
$\vartheta_{ij}=x_{ij} \frac{\partial }{ \partial x_{ij}}$. 
Then the q-oscillator superalgebra is realized by 
\begin{align}
\cd_{ij} = x_{ij}, \qquad \cc_{ij} = \frac{1}{x_{ij}} [\vartheta_{ij} ]_{q} , \qquad \n_{ij}= \vartheta_{ij} . 
 \label{q-osc-x}
\end{align}
Under this realization \eqref{q-osc-x}, 
\eqref{real1} for the distinguished grading ($p_{i}=1$ for $i\in \{1,2,\dots, M \}$, 
$p_{i}=-1$ for $i\in \{M+1,M+2,\dots, M+N \}$)
corresponds
\footnote{The formula in  \cite{AOS97} is defined  for the distinguished grading. Then we made 
a fine tune on sign factors so that the formula is valid for any gradings. 
Note that $(-1)^{p(i)p(i+1)}=p_{i}$ and 
$(-1)^{p(k)(p(i)+p(i+1))}=p_{i}p_{i+1}$ 
for $k \in \{i+2,i+3,\dots, M+N \}$ hold for the distinguished grading.
The parameters $\lambda_{i}$ and $q$ in  \cite{AOS97} correspond to 
$p_{i}\lambda_{i}- p_{i+1} \lambda_{i+1} $ and $q^{-1}$ respectively. 
The $U_{q}(sl(M|N))$ Cartan elements $h_{i}$ in  \cite{AOS97} are related to our 
$U_{q}(gl(M|N))$ Cartan elements by $h_{i}=p_{i}e_{ii} -p_{i+1}e_{i+1,i+1}$. 
The generators $e_{i}$ (resp.\ $f_{i}$) in `PROPOSITION 1. (ii)' in \cite{AOS97} 
 correspond to $ e_{i,i+1}$ (resp.\ $p_{i} e_{i+1,i}$). 
Moreover, we had to remove the term $-(\nu_{i} +\nu_{i+1}) \vartheta_{i,i+1}$ in 
the right hand side of eq.\ (18) in \cite{AOS97}, and 
put $\vartheta_{ii}=0$. 
The relation to \cite{Kimura96} can be seen from {\em Remark} 2 in \cite{AOS97}. } 
to eq.\ (25)  in \cite{AOS97}. 

By using automorphisms of the q-oscillator algebra and $U_{q}(gl(M|N))$ 
(and change of variables), 
one can derive many variants of \eqref{real1}, which superficially look different from 
the original one. 
Here we give three typical examples of them. 
First, we explain the relation between the oscillator realization \eqref{real44} 
used in the main  text and \eqref{real1}. 
Let us apply the following transformations consecutively to 
\eqref{real44}:
the rescaling of the generators of the q-oscillator algebra
\begin{align}
\begin{split}
&\cc_{ij} \mapsto 
(-1)^{\sum_{k=i+1}^{j-1}p(k)+\sum_{k=i}^{j-1}p(k)p(k+1)+p(i)p(j)}\cc_{ij},
\\[6pt]
&\cd_{ij} \mapsto 
(-1)^{\sum_{k=i+1}^{j-1}p(k)+\sum_{k=i}^{j-1}p(k)p(k+1)+p(i)p(j)}\cd_{ij}
\\[6pt]
&\n_{ij} \mapsto \n_{ij} \qquad \text{for} \quad  1 \le i<j \le M+N,
\end{split} 
\label{osrec1}
\end{align}
the automorphism of the q-oscillator algebra  
\begin{multline}
\n_{ia} \mapsto -\n_{ia}-p_{i}p_{a} ,  
\qquad 
\cc_{ia} \mapsto  \cd_{ia}, 
\qquad 
\cd_{ia} \mapsto -p_{i}p_{a} \cc_{ia} , 
%
\label{autoosc2}
\end{multline}
 the replacement 
\begin{align}
\lambda_{i} \mapsto -\lambda_{i}  +p_{i}\left(p_{[1,i-1]}-p_{[i+1,M+N]}\right) ,
\label{replacelambda}
\end{align}
and 
the automorphism of $U_{q}(gl(M|N))$
\begin{align}
e_{i,i+1} \mapsto -p_{i}p_{i+1}e_{i+1,i}, \quad 
e_{i+1,i} \mapsto -e_{i,i+1}, \quad 
e_{ii} \mapsto -e_{ii}.
\label{autogl0}
\end{align}
Then we obtain  the realization \eqref{real1}.

Let us apply the following transformations to \eqref{real1}: 
the rescaling of the q-oscillator algebra \eqref{osrec1}, 
the transformation
\footnote{The transformation \eqref{transMN} corresponds to read  
the Dynkin diagram of $gl(M|N)$ from the opposite direction. Thus 
 this effectively produces $U_{q}(gl(N|M))$ with  the opposite sign of the grading parameters. 
 In order to recover $U_{q}(gl(M|N))$, we have to change the grading parameters as in \eqref{replaceosc}. 
}
\begin{align}
e_{\alpha_{i}} \mapsto e_{\alpha_{M+N-i}}, \quad 
e_{-\alpha_{i}}  \mapsto p_{M+N-i}p_{M+N+1-i} e_{-\alpha_{M+N-i}}, \quad 
e_{ii} \mapsto -e_{M+N+1-i,M+N+1-i}
,
 \label{transMN}
\end{align}
 the  replacement
\begin{align}
\begin{split}
&
p_{i} \mapsto -p_{M+N+1-i}, \quad 
\lambda_{i} \mapsto -\lambda_{M+N+1-i}, \quad 
\n_{ia} \mapsto \n_{M+N+1-a,M+N+1-i},
\\[6pt]
 &\cc_{ia} \mapsto \cc_{M+N+1-a,M+N+1-i}, \quad 
\cd_{ia} \mapsto \cd_{M+N+1-a,M+N+1-i},
\end{split}
\label{replaceosc}
\end{align} 
and the rescaling of the generators of the q-oscillator algebra
\begin{align}
&\cc_{ij} \mapsto 
(-1)^{i-j-1}\cc_{ij},
\quad 
\cd_{ij} \mapsto 
(-1)^{i-j-1}\cd_{ij},
\quad 
\n_{ij} \mapsto \n_{ij} 
\quad \text{for} \quad  1 \le i<j \le M+N.
\label{osrec2}
\end{align}
Then  
we obtain 
\begin{align}
\begin{split}
e_{ii} & =\lambda_{i} +\n_{[1,i-1],i} - \n_{i,[i+1,M+N]} 
\quad \text{for} \quad i \in \Ik, 
\\[6pt]
e_{i,i+1} &=  \cc_{i,i+1} q^{ -p_{i}\n_{i,[i+2,M+N]} + p_{i+1}\n_{i+1,[i+2,M+N]}  } 
 \\
& \qquad 
-p_{i+1} \sum_{k=i+2}^{M+N} p_{k} \cc_{ik}  \cd_{i+1,k} 
q^{- p_{i}\n_{i,[k+1,M+N]} +p_{i+1}\n_{i+1,[k+1,M+N]} } ,
\\[6pt]
e_{i+1,i} &= 
-p_{i}\sum_{k=i+2}^{M+N} 
p_{k} \cc_{i+1,k}\cd_{ik}
\\
& \qquad \times 
 q^{-p_{i}\lambda_{i}+ p_{i+1} \lambda_{i+1} +
p_{i} \n_{i,[i+1,k-1]} - p_{i+1} \n_{i+1,[i+2,k-1]}  -
 p_{i} \n_{[1,i-1],i} + p_{i+1} \n_{[1,i],i+1}  } 
 \\
 &
 \quad 
+ p_{i}
\cd_{i,i+1} 
\left[
p_{i}\lambda_{i}- p_{i+1} \lambda_{i+1}+p_{i} \n_{[1,i-1],i} - p_{i+1} \n_{[1,i],i+1} 
\right]_{q}
 \\
& \quad 
 +
\sum_{k=1}^{i-1} 
\cd_{k,i+1} \cc_{ki} q^{p_{i}\lambda_{i}-p_{i+1}\lambda_{i+1} 
+ p_{i}\n_{[1,k],i} - p_{i+1}\n_{[1,k],i+1}  }
\quad \text{for} \quad i \in \Ik \setminus \{M+N\},
\end{split}
\label{real3}
\\
e_{j,M+N}&=
\cc_{j,M+N} 
 q^{-p_{[j+1,M+N-1]} - p_{M+N}\n_{[j+1,M+N-1],M+N}} 
 \quad \text{for} \quad j \in \Ik \setminus \{M+N\}.
 \label{enj-3}
\end{align}
Here the expression \eqref{enj-3} is obtained based on \eqref{eij}. 
Let us consider reduction of the 
q-oscillator algebra in \eqref{real3}. 
Fix parameters $a \in \{0,1,\dots, M+N\}$ and $\mu \in {\mathbb C}$, and define a set by 
$I=\{1,2,\dots,a \}$. 
We find that 
\eqref{real3} still realizes $U_{q}(gl(M|N))$ 
even if we apply the following  replacement:
\begin{align}
\cc_{ij} \mapsto 0 , 
\quad 
\cd_{ij} \mapsto 0,
\quad 
\n_{ij} \mapsto 0 ,
\quad 
\lambda_{i} \mapsto p_{i}\mu 
\quad \text{for} \quad i,j \in I. 
 \label{rel-res}
\end{align}

Let us apply the following to  \eqref{real3}:  
 the automorphisms  \eqref{autoosc2} and 
\begin{align}
\begin{split}
&\cc_{ij} \mapsto 
(-1)^{1+\sum_{k=i}^{j}p(k)+\sum_{k=i}^{j-1}p(k)p(k+1)+p(i)p(j)}\cc_{ij},
\\[6pt]
&\cd_{ij} \mapsto 
(-1)^{1+\sum_{k=i}^{j}p(k)+\sum_{k=i}^{j-1}p(k)p(k+1)+p(i)p(j)}\cd_{ij}
\\[6pt]
&\n_{ij} \mapsto \n_{ij} \qquad \text{for} \quad  1 \le i<j \le M+N,
\end{split} 
\label{osrec3}
\end{align}
of the q-oscillator algebra, the  replacement \eqref{replacelambda}, and 
the automorphism 
\begin{align}
e_{i,i+1} \mapsto -e_{i+1,i}, \quad 
e_{i+1,i} \mapsto -p_{i}p_{i+1}e_{i,i+1}, \quad 
e_{ii} \mapsto -e_{ii}, 
\label{autogl}
\end{align}
of $U_{q}(gl(M|N))$. 
We obtain  
\begin{align}
e_{ii} & =\lambda_{i} +\n_{[1,i-1],i} - \n_{i,[i+1,M+N]} 
\quad \text{for} \quad j \in \Ik , 
\nonumber  \\[6pt]
e_{i,i+1} &= 
-p_{i}
\sum_{k=i+2}^{M+N} 
p_{k}  \cc_{ik} \cd_{i+1,k}
\nonumber  \\
& \qquad \times 
 q^{p_{i}\lambda_{i}- p_{i+1} \lambda_{i+1} -
p_{i} \n_{i,[i+1,k-1]} + p_{i+1} \n_{i+1,[i+2,k-1]}  +
 p_{i} \n_{[1,i-1],i} - p_{i+1} \n_{[1,i],i+1}  } 
\nonumber  \\
 &
 \quad 
+p_{i}
\cc_{i,i+1} 
\left[
p_{i}\lambda_{i}- p_{i+1} \lambda_{i+1}+p_{i} \n_{[1,i-1],i} - p_{i+1} \n_{[1,i],i+1}
+p_{i+1}  
\right]_{q}
\nonumber  \\
& \quad 
+
\sum_{k=1}^{i-1} 
\cd_{ki} \cc_{k,i+1}  q^{-p_{i}\lambda_{i}+p_{i+1}\lambda_{i+1} 
- p_{i}\n_{[1,k],i} +p_{i+1}\n_{[1,k],i+1}-p_{i}-p_{i+1}  }
, \label{real2} \\[6pt]
e_{i+1,i} &=  \cd_{i,i+1} q^{ p_{i}\n_{i,[i+2,M+N]} - p_{i+1}\n_{i+1,[i+2,M+N]}  } 
\nonumber  \\
& \qquad 
-p_{i+1} \sum_{k=i+2}^{M+N} p_{k}   \cc_{i+1,k} \cd_{ik} 
q^{ p_{i}\n_{i,[k+1,M+N]} - p_{i+1}\n_{i+1,[k+1,M+N]} } 
\nonumber  \\
& 
\hspace{170pt} \text{for} \quad i \in \Ik \setminus \{M+N\},
\nonumber  \\[10pt]
e_{M+N,j}&=
 \cd_{j,M+N}
 q^{p_{[j+1,M+N-1]} + p_{M+N}\n_{[j+1,M+N-1],M+N}} 
\quad \text{for} \quad j \in \Ik \setminus \{M+N\}.
 \label{enj-2}
\end{align}
Here the expression \eqref{enj-2} is obtained based on \eqref{eij}. 
Let us consider reduction of the 
q-oscillator algebra in \eqref{real2}. 
Fix parameters $a \in \{0,1,\dots, M+N\}$ and $\mu \in {\mathbb C}$, and define a set by 
$I=\{1,2,\dots,a \}$. 
We find that 
\eqref{real2} still realizes $U_{q}(gl(M|N))$ 
even if we apply the following  replacement:
\begin{align}
\cc_{ij} \mapsto 0 , 
\quad 
\cd_{ij} \mapsto 0,
\quad 
\n_{ij} \mapsto 0 ,
\quad 
\lambda_{i} \mapsto p_{i}\mu 
\quad \text{for} \quad i,j \in I. 
 \label{rel-res2}
\end{align}

On the Fock space spanned by \eqref{vecFock}, 
any of \eqref{real1}, \eqref{real44}, \eqref{real3} and \eqref{real2} 
realizes a highest weight representation 
of $U_{q}(gl(M|N))$ with the hight weight 
$\lambda =(\lambda_{1},\dots , \lambda_{M+N})$ 
and the highest weight vector $ | 0 \rangle$, 
in the sense of \eqref{hwvglmn}. 
\section*{Appendix C: q-oscillator realization of contracted algebras 
in the generators $L_{ij}$ and $\Lo_{ij}$}
\addcontentsline{toc}{section}{Appendix C}
\def\theequation{C\arabic{equation}}
\setcounter{equation}{0}
%
Let us take a subset $I=\{a+1,a+2,\dots, M+N \}$ of ${\mathfrak I}$ and 
it complement set $\overline{I}={\mathfrak I} \setminus I $. 
One can rewrite 
\eqref{real44lim2} in terms of $L_{ij}$ and $\Lo_{ij}$ as follows:
\begin{align}
L_{ii} & = q^{- p_{i}\n_{i,I}}
, \quad 
\Lo_{ii}=0
 \quad \text{for} \quad   i \in \overline{I}, 
 \nonumber
\\[6pt]
L_{ii} & =q^{p_{i}(\lambda_{i} +\n_{[1,i-1],i} - \n_{i,[i+1,M+N]})}
, \quad 
\Lo_{ii} =q^{-p_{i}(\lambda_{i} +\n_{[1,i-1],i} - \n_{i,[i+1,M+N]})}
 \quad \text{for} \quad   i \in I, 
 \nonumber
\\[6pt]
L_{i+1,i} &=  
 -p_{i}p_{i+1}(q-q^{-1})
\sum_{k \in I}  p_{k}
\cc_{ik} \cd_{i+1,k} q^{
- p_{i}(\n_{i,[k,M+N]}+\n_{i,I}) +p_{i+1}\n_{i+1,[k,M+N]} +p_{i}+p_{i+1} }
\nonumber
\\
 & \qquad \text{for} \quad i,i+1 \in \overline{I},
 \nonumber
\\[6pt]
L_{i+1,i} &=  
p_{i}p_{i+1}\cc_{i,i+1} 
q^{- p_{i+1} \lambda_{i+1} - 2p_{i}\n_{i,I}+
 p_{i+1} \n_{i+1,[i+2,M+N]}  +p_{i}
}
\nonumber
 \\
& \  
 -p_{i}p_{i+1}(q-q^{-1})
\sum_{k=i+2}^{M+N}  p_{k}
\cc_{ik} \cd_{i+1,k} q^{-p_{i+1}\lambda_{i+1}
- p_{i}(\n_{i,[k,M+N]}+\n_{i,I}) +p_{i+1}\n_{i+1,[k,M+N]} +p_{i}+p_{i+1} }
\nonumber
\\
 & \qquad \text{for} \quad i \in \overline{I},  \quad i+1 \in I, \quad (i=a), 
 \nonumber
\\[6pt]
L_{i,i+1,i} &=  
p_{i+1}(q-q^{-1})
\Bigl( \ 
\sum_{k=1}^{i-1}
 \cd_{ki} \cc_{k,i+1}
 \nonumber
\\
& \quad \times 
q^{-p_{i}\lambda_{i}+ p_{i+1} \lambda_{i+1} -
p_{i} \n_{[k+1,i-1],i} + p_{i+1} \n_{[k+1,i],i+1}  +
 p_{i} \n_{i,[i+1,M+N]} - p_{i+1} \n_{i+1,[i+2,M+N]}  } 
 \label{real44lim2L}
 \\
 &
 \quad 
+
p_{i}\cc_{i,i+1} 
\left[
p_{i}\lambda_{i}- p_{i+1} \lambda_{i+1}- p_{i} \n_{i,[i+1,M+N]} +
 p_{i+1} \n_{i+1,[i+2,M+N]}  +p_{i}
\right]_{q}
\nonumber
 \\
& \quad 
 -p_{i}
\sum_{k=i+2}^{M+N}  p_{k}
\cc_{ik} \cd_{i+1,k} q^{p_{i}\lambda_{i}-p_{i+1}\lambda_{i+1} 
- p_{i}\n_{i,[k,M+N]} +p_{i+1}\n_{i+1,[k,M+N]} +p_{i}+p_{i+1} }
\Bigr)
\nonumber 
\\
&\quad 
\times 
q^{p_{i}(\lambda_{i} +\n_{[1,i-1],i} - \n_{i,[i+1,M+N]})} 
 \qquad \text{for} \quad i,i+1 \in I,
 \nonumber
 \\[6pt]
\Lo_{ji} &= 0 \quad \text{for} \quad i,j \in \overline{I}, \quad  i>j,
\nonumber
\\[6pt]
\Lo_{i,i+1} &=-p_{i}(q-q^{-1})  \cd_{i,i+1} q^{p_{i}(1+\n_{i,I})  - p_{i+1}\n_{[1,i-1],i+1}  } 
 \quad \text{for} \quad i \in \overline{I}, \quad i+1 \in I, \quad (i=a),
 \nonumber
\\[6pt]
\Lo_{i,i+1} &=   -p_{i}(q-q^{-1}) \Bigl(
\cd_{i,i+1} q^{ p_{i}\n_{i,[i+1,M+N]} - p_{i+1}\n_{[1,i-1],i+1}  } 
\nonumber
\\
&\ +
\sum_{k=1}^{i-1}  \cd_{k,i+1} \cc_{ki} q^{- p_{i}(\n_{[k,i-1],i}-\n_{i,[i+1,M+N]}) - p_{i+1}\n_{[1,k-1],i+1} } 
\Bigr)q^{p_{i}(1-\lambda_{i})}
\quad \text{for} \quad i,i+1 \in I,
 \nonumber
\\[6pt]
\Lo_{1i}&= -p_{1}(q-q^{-1}) \cd_{1i}  q^{ p_{1}(1+\n_{1,[i,M+N]})} 
 \quad \text{for} \quad  
 i \in I,  \quad i>1.
\nonumber
\end{align}
One can rewrite 
\eqref{real44lim3} in terms of $L_{ij}$ and $\Lo_{ij}$ as follows:
\begin{align}
L_{ii} & = q^{- p_{i}\n_{i,I}}
 , \qquad 
\Lo_{ii}=0
 \quad \text{for} \quad   i \in \overline{I}, 
\nonumber \\[6pt] 
L_{ii} &=q^{\mu+p_{i}\n_{\overline{I},i}}
, \qquad 
\Lo_{ii} =q^{-\mu-p_{i}\n_{\overline{I},i}}
 \quad \text{for} \quad   i \in I, 
\nonumber  \\[6pt]
L_{i+1,i} &=  
 -p_{i}p_{i+1}(q-q^{-1})
\sum_{k \in I}  p_{k}
\cc_{ik} \cd_{i+1,k} q^{
- p_{i}(\n_{i,[k,M+N]}+\n_{i,I}) +p_{i+1}\n_{i+1,[k,M+N]} +p_{i}+p_{i+1} }
\nonumber  \\
 & \qquad \text{for} \quad i,i+1 \in \overline{I},
\nonumber  \\[6pt]
L_{i+1,i} &=  
p_{i}p_{i+1}\cc_{i,i+1} 
q^{-\mu - 2p_{i} \n_{i,I}  +p_{i}
}
%
  \quad \text{for} \quad i \in \overline{I},  \quad i+1 \in I, \quad (i=a), 
\nonumber  \\[6pt]
L_{i+1,i} &=p_{i+1}(q-q^{-1})  
\sum_{k \in \overline{I}}
 \cd_{ki} \cc_{k,i+1}
q^{\mu +
p_{i} \n_{[1,k],i} + p_{i+1} \n_{[k+1,a],i+1}   } 
 \quad \text{for} \quad i,i+1 \in I,
\label{real44lim3L}  \\[6pt]
\Lo_{ji} &= 0 \quad \text{for} \quad i,j \in \overline{I}, \quad  i>j,
\nonumber  \\[6pt]
\Lo_{i,i+1} &= -p_{i}(q-q^{-1}) \cd_{i,i+1} q^{p_{i}(1+\n_{i,I})  - p_{i+1}\n_{[1,i-1],i+1}  } 
 \quad \text{for} \quad i \in \overline{I}, \quad i+1 \in I, \quad (i=a),
\nonumber  \\[6pt]
\Lo_{i,i+1} &=  - p_{i}(q-q^{-1})
\sum_{k \in \overline{I}}  
\cd_{k,i+1} \cc_{ki} q^{-\mu+ p_{i}(1-\n_{[k,a],i}) - p_{i+1}\n_{[1,k-1],i+1} } 
 \quad \text{for} \quad i,i+1 \in I,
\nonumber  \\[6pt]
\Lo_{1i}&= -p_{1}(q-q^{-1})\cd_{1i}  q^{ p_{1}(1+\n_{1,[i,M+N]})} 
 \quad \text{for} \quad
  i \in I,  \quad 1 \in \overline{I},  \quad i>1.
\nonumber  
\end{align}

\section*{Appendix D: q-Holstein-Primakoff realization and L-operators 
for Baxter Q-operators 
(supplement for \cite{Tsuboi12})}
\addcontentsline{toc}{section}{Appendix D}
\def\theequation{D\arabic{equation}}
\setcounter{equation}{0}
In this section, we will rederive the L-operators for Q-operators proposed in \cite{Tsuboi12}, 
which are degenerated solutions of the graded Yang-Baxter equation \eqref{YB-L}, 
 by taking limits of a q-analogue of the Holstein-Primakoff realization of $U_{q}(gl(M|N))$. 
\subsection*{q-Holstein-Primakoff realization of $U_{q}(gl(M|N))$}
Take an element $i \in {\mathfrak I}$, and 
define $I=\{ i \}$, $\overline{I}={\mathfrak I} \setminus \{ i \}$ 
(we assume that $i $ is a constant number throughout this section). 
In the main text, the generators 
$\{ \cc_{\alpha \beta}, \cd_{\alpha \beta}, \n_{\alpha \beta} \}$ 
of the q-oscillator algebra are defined for $\alpha, \beta \in {\mathfrak I} $, $\alpha < \beta$. 
In this section, we change this to $(\alpha, \beta) \in I \times \overline{I}$ 
(the parities of the generators and the relations are defined in the same manner).  
Then we define
\begin{align} 
& e_{ii}= p_{i}m -  \n_{i,\overline{I}}, 
\label{e-HP-1} 
\\
&e_{aa}= \n_{ia} 
\qquad \text{for} \quad a \in \overline{I}, \\
&e_{ia}=(q-q^{-1})^{-1} \cc_{ia} q^{p_{i} ( \n_{i,[i+1,a-1]} +\n_{i,\overline{I}}) }
 \qquad \text{for} \quad i+1 \le a \le M+N, \\
&e_{bi}=-p_{i}(q-q^{-1}) \cd_{ib} 
  \left[ m- p_{i} \n_{i, \overline{I}}  \right]_{q} 
    \nonumber \\
 &  \qquad  \qquad \times     q^{m- p_{i} (  \n_{i,[1,b-1]} +\n_{i,[i+1,M+N]} ) 
  -p_{b}  \n_{ib}} 
\qquad \text{for} \quad 1 \le b \le i-1, \\
&e_{ba}= \cd_{ib} \cc_{ia} q^{p_{i}  \n_{i,[b,a-1]}  -p_{b}  \n_{ib} } 
 \nonumber \\
 & \qquad \qquad \text{for} \quad 1 \le b<a \le i-1
\quad \text{or} \quad i+1 \le b<a \le M+N,  \\
&e_{ba}= - \cd_{ib}  \cc_{ia}
  q^{2m +p_{i}(1- \n_{i,[1,b-1]} -\n_{i,[a,M+N]} )   -p_{b}  \n_{ib} } 
 \nonumber \\
 & \qquad \qquad \text{for} \quad 1 \le b<i<a \le  M+N,
\\[8pt]
&e_{ba}=\cd_{ib} \cc_{ia}
  q^{p_{i}(1- \n_{i,[a,b-1]} ) - p_{a}(1- \n_{ia} ) } 
 \nonumber \\
 & \qquad \qquad \text{for} \quad 1 \le a<b \le i-1 
 \quad \text{or} \quad i+1 \le a < b  \le  M+N,
 \\
&e_{ia}= - (q-q^{-1})^{-1}\cc_{ia} q^{-m+ p_{i} 
  (\n_{i,[1,a-1]}+\n_{i,[i+1,M+N]}) +p_{a}(\n_{ia}-1) }
\nonumber \\
& \qquad  \qquad \text{for} \quad 1 \le a \le i-1, \\
&e_{bi}=p_{i}(q-q^{-1}) \cd_{ib} 
 \left[ m - p_{i} \n_{i, \overline{I}} \right]_{q}
  q^{- p_{i}( 1+ \n_{i,[i+1,b-1]} +\n_{i,\overline{I}} )}
  \nonumber \\
  & \qquad  \qquad \text{for} \quad i+1 \le b \le M+N, \\
& e_{ba}= - \cd_{ib} \cc_{ia}
  q^{-2m+ p_{i}  (\n_{i,[1,a-1]}+\n_{i,[b,M+N]}) -p_{a}(1-\n_{ia}) } 
 \nonumber \\
 & 
\qquad \qquad \text{for} \quad 1 \le a <i <  b  \le M+N,
 \label{e-HP-10}
\end{align} 
where $m \in {\mathbb C}$.
This is a q-analogue of the Holstein-Primakoff realization of $U_{q}(gl(M|N))$ 
(cf.\ \cite{P98}).
For $I=\{1\}$, this realizes 
an infinite dimensional representation 
with the highest weight $\lambda =(p_{1}m,0,\dots , 0)$ 
and the highest weight vector $|  0  \rangle$ on the Fock space 
in the sense of \eqref{hwvglmn}. 
However, this is not the case for $I=\{i\}$, $i\ne 1$. The vacuum vector 
 $|  0  \rangle$ 
carries the weight (eigenvalue of $e_{aa}$)
$\lambda_{a} =p_{a}m\delta_{ia}$ ($1\le a \le M+N$) and is killed 
at least by 
$e_{ab}$ for $1 \le a<b <i$, $i \le a<b  \le M+N$ and $1 \le b <i \le a \le M+N$.
 
 Under the reduction \eqref{rel-res1} for $I=\{2,3,\dots,M+N\}$,  \eqref{real1} and
  \eqref{e1j}   
(and $e_{jk}$ from \eqref{eij}) 
for $\lambda_{j}=p_{1}m\delta_{j1}$ and $\mu=0$ 
coincides with \eqref{e-HP-1}-\eqref{e-HP-10} for 
$I=\{1\}$ if the following automorphism of 
the q-oscillator algebra is applied to \eqref{e-HP-1}-\eqref{e-HP-10}. 
\begin{align}
\begin{split}
&
\n_{1a} \to \n_{1a} 
\quad 
\cc_{1a} \to (q-q^{-1}) \cc_{1a}q^{-p_{1} \n_{1,\overline{I}}},
\quad 
\cd_{1a} \to (q-q^{-1})^{-1} q^{p_{1} \n_{1,\overline{I}}}  \cd_{1a} 
\quad \text{for} 
\  a \in \overline{I} .
\end{split}
\end{align}
We remark that 
the notation $I$ and $\overline{I}$ have to be exchanged for comparison 
between \eqref{real1}-\eqref{e1j} and \eqref{e-HP-1}-\eqref{e-HP-10}. 
\subsection*{L-operator}
Plugging \eqref{e-HP-1}-\eqref{e-HP-10} into the formula \eqref{maprll1}-\eqref{maprll3}, 
we obtain the following elements of an L-operator. 
\begin{align}
& L_{\alpha \beta}=0 \quad \text{for} 
\quad \alpha < \beta  ,  
\label{L-HP-1} \\
& L_{ii}= q^{m - p_{i} \n_{i,\overline{I}}}, 
\\
&L_{aa}=q^{p_{a}  \n_{ia}} 
\qquad \text{for} \quad a \in \overline{I}, \\
&L_{ai}=p_{a}\cc_{ia} q^{m+p_{i}  \n_{i,[i+1,a-1]}}
 \qquad \text{for} \quad i+1 \le a \le M+N, \\
&L_{ib}=-(q-q^{-1})^{2} \cd_{ib} 
  \left[ m- p_{i} \n_{i, \overline{I}}  \right]_{q} 
    q^{m- p_{i} (  \n_{i,[1,b-1]} +\n_{i,[i+1,M+N]} ) } 
    \nonumber \\
 &  \qquad \qquad \text{for} \quad 1 \le b \le i-1, \\
&L_{ab}=p_{a}
 (q-q^{-1}) \cd_{ib} \cc_{ia}
  q^{p_{i}  \n_{i,[b,a-1]}} 
 \nonumber \\
 & \qquad \qquad \text{for} \quad 1 \le b<a \le i-1
\quad \text{or} \quad i+1 \le b<a \le M+N,  \\
&L_{ab}=-p_{a}
 (q-q^{-1})\cd_{ib}  \cc_{ia}
  q^{2m +p_{i}(1- \n_{i,[1,b-1]} -\n_{i,[a,M+N]} ) } 
 \nonumber \\
 & \qquad \qquad \text{for} \quad 1 \le b<i<a \le  M+N,
\\[8pt]
& \Lo_{\alpha \beta}=0 \quad \text{for} 
\quad \alpha > \beta 
, \\
& \Lo_{ii}= q^{-m+p_{i} \n_{i,\overline{I}}}, \\
&\Lo_{aa}=q^{-p_{a}  \n_{ia}} 
\qquad \text{for} \quad a \in \overline{I}, \\
&\Lo_{ab}=-p_{a}
 (q-q^{-1}) \cd_{ib} \cc_{ia}
  q^{p_{i}(1- \n_{i,[a,b-1]} ) } 
 \nonumber \\
 & \qquad \qquad \text{for} \quad 1 \le a<b \le i-1 
 \quad \text{or} \quad i+1 \le a < b  \le  M+N,
 \\
&\Lo_{ai}=p_{a}\cc_{ia} q^{-m+ p_{i} 
  (\n_{i,[1,a-1]}+\n_{i,[i+1,M+N]})}
 \qquad \text{for} \quad 1 \le a \le i-1, \\
&\Lo_{ib}=-(q-q^{-1})^{2} \cd_{ib} 
 \left[ m - p_{i} \n_{i, \overline{I}} \right]_{q}
  q^{-m- p_{i} \n_{i,[i+1,b-1]} }
  \nonumber \\
  & \qquad  \qquad \text{for} \quad i+1 \le b \le M+N, \\
& \Lo_{ab}=p_{a}
 (q-q^{-1}) \cd_{ib} \cc_{ia}
  q^{-2m+ p_{i}  (\n_{i,[1,a-1]}+\n_{i,[b,M+N]})} 
 \nonumber \\
 & 
\qquad \qquad \text{for} \quad 1 \le a <i <  b  \le M+N,
 \label{L-HP-14}
\end{align} 
where $i \in I$.
\subsection*{Limit of the L-operator: $q^{m} \to 0$ case}
After making a shift $m\to m+p_{i}\mu $ in \eqref{L-HP-1}-\eqref{L-HP-14}, 
we consider 
 a renormalized L-operator [see \ eq.\ (3.79) in \cite{Tsuboi12} 
for $\mu=0$ case]:
\begin{align}
\tilde{\mathbf L}(x)={\mathbf L}(xq^{-2m})(1 \otimes q^{-m \sum_{j \in I} E_{jj}}) .
 \label{renoL}
\end{align}
In components, this is transcribed as
\begin{align}
\tilde{L}_{jk}=L_{jk}q^{-m\delta_{ki}}, 
\qquad 
\tilde{\Lo}_{jk}=\Lo_{jk}q^{m(2- \delta_{ki})} 
\qquad \text{for} \quad j,k \in {\mathfrak I}, 
\quad i \in I.
\end{align}
Then we find that the components of  
  the L-operator $\Lf^{-}(x)=\lim_{ q^{m} \to 0}\tilde{\mathbf L}(x)$ 
 are given
 \footnote{Here $q$ is assumed to be a constant number. 
 The limits of $\tilde{L}_{jk}$ and $\tilde{\Lo}_{jk}$ are 
 denoted again as $L_{jk}$ and $ \Lo_{jk}$.}
  by 
\begin{align}
& L_{\alpha \beta}=0 \quad \text{for} 
\quad \alpha < \beta   \quad \text{or} \quad 1 \le \beta <i<\alpha  \le  M+N ,  
\label{L-HP-1-1} \\
& L_{ii}= q^{p_{i}\mu - p_{i} \n_{i,\overline{I}}}, 
\\
&L_{aa}=q^{p_{a}  \n_{ia}} 
\qquad \text{for} \quad a \in \overline{I}, \\
&L_{ai}=p_{a}\cc_{ia} q^{p_{i}\mu +p_{i}  \n_{i,[i+1,a-1]}}
 \qquad \text{for} \quad i+1 \le a \le M+N, \\
&L_{ib}=(q-q^{-1}) \cd_{ib} 
    q^{p_{i}  \n_{i,[b,i-1]} } 
  \qquad \text{for} \quad 1 \le b \le i-1, \\
&L_{ab}=p_{a}
 (q-q^{-1}) \cd_{ib} \cc_{ia}
  q^{p_{i}  \n_{i,[b,a-1]}} 
 \nonumber \\
 & \qquad \qquad \text{for} \quad 1 \le b<a \le i-1
\quad \text{or} \quad i+1 \le b<a \le M+N, 
\\[8pt]
& \Lo_{\alpha \beta}=0 \quad \text{for} 
\quad \alpha > \beta ,
 \quad 1 \le \alpha \le \beta \le i-1 
 \quad \text{or} \quad i+1 \le \alpha \le  \beta  \le  M+N,
 \\
& \Lo_{ii}= q^{-p_{i}\mu+p_{i} \n_{i,\overline{I}}}, 
 \\
&\Lo_{ai}=p_{a}\cc_{ia} q^{-p_{i}\mu+ p_{i} 
  (\n_{i,[1,a-1]}+\n_{i,[i+1,M+N]})}
 \qquad \text{for} \quad 1 \le a \le i-1, \\
&\Lo_{ib}=(q-q^{-1}) \cd_{ib} 
  q^{-2p_{i}\mu +p_{i} (\n_{i,[1,i-1]}+\n_{i,[b,M+N]}) }
   \qquad \text{for} \quad i+1 \le b \le M+N, \\
& \Lo_{ab}=p_{a}
 (q-q^{-1}) \cd_{ib} \cc_{ia}
  q^{-2p_{i}\mu+ p_{i}  (\n_{i,[1,a-1]}+\n_{i,[b,M+N]})} 
 \nonumber \\
 & 
\qquad \qquad \text{for} \quad 1 \le a <i <  b  \le M+N,  
 \label{L-HP-1-14}
\end{align} 
where $i \in I$.
These equations \eqref{L-HP-1-1}-\eqref{L-HP-1-14} for $\mu=0$ 
 precisely coincide
 \footnote{
 From \eqref{L-HP-1-1}-\eqref{L-HP-1-14} for $\mu=0$, $i=M+N$, $N=0$, 
 one can also reproduce the q-oscillator representation of the
  Borel subalgebra ${\mathcal B}_{+}$ of $U_{q}(\hat{sl}(M))$
 for Baxter Q-operators found in \cite{Kojima08}. 
 Substituting \eqref{L-HP-1-1}-\eqref{L-HP-1-14} for $\mu=0$, $i=M+N$, $N=0$ 
 into eq. (3.82) in \cite{Tsuboi12}, one obtains 
 $e_{j}=\cd_{M,j}\cc_{M,j+1}$, 
 $h_{j}=\n_{M,j}-\n_{M,j+1}$ for $1\le j \le M-2$,
 $e_{M-1}=\cd_{M,M-1}$, 
 $h_{M-1}=\n_{M,M-1}+\n_{M,\overline{I}}$, 
 $e_{M}=-x(q-q^{-1})^{-1}\cc_{M,1}q^{\n_{M,\overline{I}}}$,
 $h_{M}=-\n_{M,\overline{I}}-\n_{M,1}$.  
 Set $\cc_{M,j}=-(q-q^{-1})\varepsilon_{j}q^{{\mathcal H}_{j}}$, 
 $\cd_{M,j}=\varepsilon_{j}^{*}$, 
 $\n_{M,j}=-{\mathcal H}_{j}$ for $1 \le j \le M-1$, and apply 
 the automorphism  of ${\mathcal B}_{+}$: 
 $e_{1} \mapsto tq^{\frac{1}{2}}e_{1}$,  
 $e_{j} \mapsto q^{\frac{1}{2}}e_{j}$ 
 for $2 \le j \le M-2$,  $e_{M-1} \mapsto e_{M-1}$, 
 $e_{M} \mapsto x^{-1}e_{M}$, 
 $h_{j} \mapsto h_{j}$ for $1\le j \le M$ 
 ($\varepsilon^{*}_{j},\varepsilon_{j},{\mathcal H}_{j},t$ 
 are symbols in \cite{Kojima08}). 
 Then one obtains eq. (2.2) in \cite{Kojima08} after 
 the transformation  $q \to q^{-1}$.}
  with a q-oscillator solution of the graded Yang-Baxter equation
  found in  \cite{Tsuboi12} 
[eqs.\ (3.49)-(3.59) in \cite{Tsuboi12}]. 

Let us apply the automorphism 
\begin{align}
\cc_{ia} \to q^{-p_{i}\mu} \cc_{ia}, \qquad 
\cd_{ia} \to q^{p_{i}\mu} \cd_{ia} 
\qquad \text{for} \quad i+1 \le a \le M+N , \quad i \in I
\end{align}
of the q-oscillator algebra to  \eqref{L-HP-1-1}-\eqref{L-HP-1-14} 
and consider 
\begin{align}
{\mathbf L}^{- \prime }(x)={\mathbf L}^{-}(xq^{-p_{i}\mu}) .
 \label{renoL1p}
\end{align}
The components $L_{jk}$ and  $\Lo_{jk}$ of   $\Lf^{\prime}$ and $\Lbf^{\prime} $ in 
this renormalized L-operator ${\mathbf L}^{- \prime }(x)=\Lf^{\prime}-x^{-1}\Lbf^{\prime}$
 do not depend on the parameter $\mu$ except 
for the element $L_{ii}$. It satisfies  
$L_{ii}\overline{L}_{ii}=\overline{L}_{ii}L_{ii}=q^{p_{i}\mu}$ 
for $i \in I$ 
instead of  \eqref{red1}. 
We remark that components of $\Lf^{\prime}$ and $\Lbf^{\prime} $  
 realize a more degenerate algebra than $U_{q}(gl(M|N;I))$ in the limit 
$q^{p_{i}\mu} \to 0$. 
In fact, they satisfy a condition $L_{ii}=0$ for $i \in I$ in addition to \eqref{red2}. 
A twisted version of such an  
L-operator (in the sense of \cite{Reshetikhin:1990ep}) for $N=0$ case was used to construct 
a matrix product formula for symmetric Macdonald polynomials \cite{CGW15} 
(see \cite{WZ16} for related L-operators for $M+N \le 3$). 
The same type of L-operators also appeared in the context of 
quantization of soliton cellular automata \cite{IKO04}. 

\subsection*{Limit of the L-operator: $q^{m} \to \infty$ case}
We can consider the opposite limit ($q^{m} \to \infty$) for another renormalized L-operator
\begin{align}
\tilde{\tilde{\mathbf L}}(x)={\mathbf L}(x)(1 \otimes q^{-m \sum_{j \in I} E_{jj}})  .
 \label{renoL2}
\end{align}
After applying an automorphism 
\begin{align}
\cc_{ia} \to q^{2m} \cc_{ia}, \qquad 
\cd_{ia} \to q^{-2m} \cd_{ia} 
\qquad \text{for} \quad 1 \le a \le i-1 , 
\quad i \in I
\end{align}
of the q-oscillator algebra to  \eqref{L-HP-1}-\eqref{L-HP-14} 
and plugging them into \eqref{renoL2}, we  take  
 the limit 
\footnote{
The components of $\Lf$ and $\Lbf$ in $ \Lf^{+}(x)=\Lf-x^{-1}\Lbf$ 
are denoted as $L_{jk} $ and $\Lo_{jk}$ respectively.}  
 $ \Lf^{+}(x)=\lim_{q^{m}\to \infty} \tilde{\tilde{\mathbf L}}(x)$
 to get
\begin{align}
& L_{\alpha \beta}=0 \quad \text{for} 
\quad \alpha < \beta  ,  
\label{L-HP-2-1} \\
& L_{ii}= q^{ - p_{i} \n_{i,\overline{I}}}, 
\\
&L_{aa}=q^{p_{a}  \n_{ia}} 
\qquad \text{for} \quad a \in \overline{I}, \\
&L_{ai}=p_{a}\cc_{ia} q^{p_{i}  \n_{i,[i+1,a-1]}}
 \qquad \text{for} \quad i+1 \le a \le M+N, \\
&L_{ib}=-(q-q^{-1}) \cd_{ib} 
    q^{- p_{i} ( \n_{i, \overline{I}}+ \n_{i,[1,b-1]} +\n_{i,[i+1,M+N]} ) } 
 \qquad \text{for} \quad 1 \le b \le i-1, \\
&L_{ab}=p_{a}
 (q-q^{-1}) \cd_{ib} \cc_{ia}
  q^{p_{i}  \n_{i,[b,a-1]}} 
 \nonumber \\
 & \qquad \qquad \text{for} \quad 1 \le b<a \le i-1
\quad \text{or} \quad i+1 \le b<a \le M+N,  \\
&L_{ab}=-p_{a}
 (q-q^{-1})\cd_{ib}  \cc_{ia}
  q^{p_{i}(1- \n_{i,[1,b-1]} -\n_{i,[a,M+N]} ) } 
 \nonumber \\
 & \qquad \qquad \text{for} \quad 1 \le b<i<a \le  M+N,
\\[8pt]
& \Lo_{\alpha \beta}=0 \quad \text{for} 
\quad \alpha > \beta 
, \\
& \Lo_{ii}= 0, \\
&\Lo_{aa}=q^{-p_{a}  \n_{ia}} 
\qquad \text{for} \quad a \in \overline{I}, \\
&\Lo_{ab}=-p_{a}
 (q-q^{-1}) \cd_{ib} \cc_{ia}
  q^{p_{i}(1- \n_{i,[a,b-1]} ) } 
 \nonumber \\
 & \qquad \qquad \text{for} \quad 1 \le a<b \le i-1 
 \quad \text{or} \quad i+1 \le a < b  \le  M+N,
 \\
&\Lo_{ai}=p_{a}\cc_{ia} q^{ p_{i} 
  (\n_{i,[1,a-1]}+\n_{i,[i+1,M+N]})}
 \qquad \text{for} \quad 1 \le a \le i-1, \\
&\Lo_{ib}=-(q-q^{-1}) \cd_{ib} 
  q^{- p_{i}(\n_{i, \overline{I}}+ \n_{i,[i+1,b-1]}) }
 \qquad \text{for} \quad i+1 \le b \le M+N, \\
& \Lo_{ab}=p_{a}
 (q-q^{-1}) \cd_{ib} \cc_{ia}
  q^{ p_{i}  (\n_{i,[1,a-1]}+\n_{i,[b,M+N]})} 
 \nonumber \\
 & 
\qquad \qquad \text{for} \quad 1 \le a <i <  b  \le M+N,
 \label{L-HP-2-14}
\end{align} 
 where $i \in I$.
We consider two kinds of 
automorphisms of 
the q-oscillator algebra \eqref{qosc}:
\eqref{autoosc2} and 
\begin{align}
& 
\n_{ia} \mapsto \n_{ia},    \qquad \text{for}\quad  a \in \overline{I} ,
\nonumber 
\\[5pt]
&
\cc_{ia} \mapsto
p_{a}p_{i} (q-q^{-1})^{-1} \cc_{ia} 
  q^{- p_{i} (\n_{i,\overline{I}} -\n_{ia} ) 
   -p_{[1,a-1] -p_{[i,M+N]}}
   }, 
\nonumber 
\\[8pt]
&
\cd_{ia} \mapsto
p_{a}p_{i} (q-q^{-1})
  q^{p_{i} (\n_{i,\overline{I}} -\n_{ia} ) 
   +p_{[1,a-1] +p_{[i,M+N]}}
   } \cd_{ia} 
  \quad \text{for} \quad 1 \le a \le i-1, 
\nonumber 
\\[10pt]
&
\cc_{ia} \mapsto
p_{a}p_{i} (q-q^{-1})^{-1} \cc_{ia} 
  q^{- p_{i} (\n_{i,\overline{I}} -\n_{ia} ) 
   -p_{[i+1,a-1] +p_{i}}
   }, 
\nonumber 
\\[5pt]
&
\cd_{ia} \mapsto
p_{a}p_{i} (q-q^{-1})
  q^{p_{i} (\n_{i,\overline{I}} -\n_{ia} ) 
   +p_{[i+1,a-1] -p_{i}}
   } \cd_{ia} 
  \quad \text{for} \quad i+1 \le a \le M+N. 
 \label{autoosc}
\end{align}
Let us apply the automorphism \eqref{autoosc2} to 
\eqref{L-HP-2-1}-\eqref{L-HP-2-14} first, and then 
\eqref{autoosc}. We find  that 
 the renormalized L-operator
\begin{align}
{\mathbf L}^{+ \prime}(x)=
 (1 \otimes q^{p_{i} \sum_{b \in \overline{I} } E_{bb} -p_{\overline{I}}E_{ii}} )
  {\mathbf L}^{+}(xq^{2p_{i}})  .
 \label{renoL3}
\end{align}
 precisely coincides
 \footnote{We have to swap the notation 
 $\{I,\n_{\alpha \beta},\cd_{\alpha \beta} \}$ and 
 $\{\overline{I},\n_{\beta \alpha},\cd_{\beta \alpha} \}$  
 to make comparison.} 
 with another q-oscillator solution of the graded Yang-Baxter equation
  found in  \cite{Tsuboi12} 
[eqs.\ (3.60)-(3.72) in \cite{Tsuboi12}]. 

\end{document}